\documentclass[12pt,aps,showpacs]{revtex4}
\usepackage{amssymb,amsmath}
\usepackage{graphicx}
\begin{document}
\title{Method for implementation of universal quantum logic gates
in a scalable Ising spin quantum computer}
\author{G. P. Berman$^1$, D. I. Kamenev$^1$, R. B. Kassman$^{1,2}$,
C. Pineda$^3$, and V. I. Tsifrinovich$^4$}
\affiliation{$^1$ Theoretical Division and Center for Nonlinear Studies,
Los Alamos National Laboratory, Los Alamos, New Mexico 87545}
\affiliation{$^2$ Department of Physics, University of Illinois at
Urbana-Champaign, Urbana, Illinois 61801}

\affiliation{$^3$ Universidad Nacional Aut\'onoma de M\'exico, Apdo.
Postal 20-364, Mexico D. F. 01000, Mexico}
\affiliation{$^4$IDS Department, Polytechnic University,
Six Metrotech Center, Brooklyn, New York 11201}

\vspace{3mm}
\begin{abstract}

We present protocols for implementation of
universal quantum gates on an {\it arbitrary superposition} of
quantum states
in a scalable solid-state Ising spin quantum computer.
The spin chain is composed of identical spins $1/2$ with
the Ising interaction between the neighboring spins. The selective
excitations of the spins
are provided by the gradient of the external magnetic field.
The protocols are built of rectangular radio-frequency pulses.
The phase and probability errors caused by unwanted
transitions are minimized and computed numerically.
\end{abstract}
\pacs{03.67.Lx,~75.10.Jm}
\maketitle

\section{Introduction}
Current proposals on a scalable solid-state spin quantum computer can
be roughly divided into two main streams. The first one relies on a
controllable interaction between the electron spins (see, for
example, \cite{1,2}). In such proposals a point contact gate switches
interaction between the neighboring qubits. It was shown, in particular,
that the switchable Heisenberg interaction alone can provide
the universal quantum computation. The main advantage of the controllable
interaction proposals is the high clock speed in the GHz range. The
main technical challenge in such proposals is the precise control of
the interaction. It was estimated that the interaction constant must
be controllable with the accuracy $10^{-4}$. Also the time of turning
on (off) the interaction must be about 10-100 fs. The other challenge
is a possible decoherence caused by the point contact.

The second stream relies on the permanent interaction between spins
(see, for example \cite{3,3a}). In such proposals electromagnetic
pulses provide implementation of quantum computation. The main
advantage of these proposals is the use of well developed technique
of electromagnetic pulses, and the absence of the electro-static gates.
The main technical challenge for these proposals is the creation of a large gradient of the magnetic field. The other
challenge is the decoherence caused by the sources of the magnetic
field.
The latest development of the micropattern wires technique provides the
magnetic field gradients $10^5-10^6$ T/m \cite{4,5}. Even greater
gradients $\sim 10^7$ T/m are expected if the micropattern wires
will be replaced by the ferromagnetic plates \cite{6}. This development
provides new ground for the electromagnetic pulse proposals.
The field gradient $10^6$ T/m corresponds to the magnetic field
difference $0.03$ T for the distance 30 nm. In turn,
the difference between the electron spin resonance (ESR) frequencies
$\Delta\omega/2\pi$ for this distance will be 840 MHz.

For the electromagnetic pulse proposals the most convenient interaction
between qubits is, definitely, the Ising interaction \cite{6a,6b}.
Normally, the Ising interaction is considered as non-typical for solids.
However, if the frequency difference between the
neighboring spins is greater than the spin-spin interaction, the
spin-spin interaction (Heisenberg or dipole-dipole) becomes
effectively the Ising interaction. This effect is well-known in
liquid nuclear magnetic resonance (NMR) where the scalar coupling,
which is equivalent to the Heisenberg interaction, becomes
the Ising interaction due to the chemical shift of the NMR
frequencies \cite{7,8}. The same effect takes place for the
heteronuclear dipole-dipole interaction in solids. We should note,
that the long-range dipole-dipole interaction in a spin chain
can be effectively suppressed if the angle between the chain and
the external magnetic field equals to the magic
angle \cite{7,9}.

We believe that the large field gradients provide the resurgence
of the interest to the Ising spin quantum computer. If the
interaction between the
neighboring paramagnetic spins is of the order of 10MHz and the ESR
frequency difference is $\Delta\omega/2\pi=840$ MHz, the spin-spin interaction
becomes effectively the Ising interaction. As an example, we consider
phosphorus impurity donors in silicon. In the first approximation, the
exchange interaction constant for the donor electrons separated by the
distance $r$ is given by~\cite{Kane}
$$
J(r)=0.8{e^2\over\epsilon a_0}\left(\frac ra_0\right)^{5/2}\exp(-2r/a_0).
$$
Here $\epsilon$ is the dielectric constant of silicon, $a_0$ is the
effective radius of the impurity atom, and $e$ is the electron charge.
Taking $\epsilon=12$, $a_0=3$ nm, and $r=30$ nm, we obtain $J=5$~MHz.

Currently, there are three main approaches for
implementation of quantum
logic gates using the electromagnetic pulses. The first approach
is developed
in experiments with liquid state NMR quantum computer
(see, for, example, \cite{7}). To provide the conditional logic
this approach takes advantage of the spin-spin
interaction during the time-interval between the electromagnetic
pulses. Thus, the Rabi frequency $\Omega$ of the pulses
[$\Omega=\gamma B_1$, where $\gamma$ is the gyromagnetic ratio and
$B_1$ is the amplitude of the radio frequency ({\it rf}) field] is much
greater than the Ising interaction constant $J$, but the
clock speed of the gate is $J$.

The second approach is intended to
increase the clock speed by application
of more powerful strongly modulated pulses which average to zero all
undesired interactions between spins but drive the desired quantum
transitions. The first experiments indicate the almost tenfold increase
of the clock speed \cite{10}.

The third approach is based on the application of the selective
electromagnetic pulses with the Rabi frequency $\Omega<J$
(see, for example, \cite{9,BI1,BI2}). These pulses drive a resonant spin
depending on the states of its neighbors. The clock speed in this
approach is determined by the Rabi frequency, $\Omega$.
While the third approach has a clear disadvantage in the clock speed, its
advantage is the
small power of the pulses: for a given angle of the spin rotation
the power of the pulse is inversely proportional to its duration.
The small power of the pulses is not important for the room
temperature NMR quantum computer but can be important for a
scalable solid-state quantum computer, which is expected to operate
at low temperature. Besides, the powerful pulses can contribute to the
decoherence rate.

Based on the above remarks, we believe that the development of
the theoretical background for
the quantum logic gate implementation using the selective
electromagnetic
pulses in the Ising spin quantum computer is an important
task. In this paper, we present a scheme for implementation of
universal quantum gates on an arbitrary superposition of quantum states
in the Ising spin quantum computer.

The paper is organized as follows. The general problem to be solved
is formulated in Sec.~II. The Ising spin quantum computer model is
described in Sec.~III. The probability errors are
minimized in Secs.~IV and V. The phase errors are minimized in
Secs.~VI and VII. The universal gates are tested numerically
in Sec.~VIII
by the exact numerical modeling of quantum dynamics of the system.
In Sec. IX we summarize our results.

\section{Formulation of the problem}
The most important quantum algorithms exploit the property of
quantum interference. In the result of implementation of these
algorithms a small amount of ``useful'' states is amplified
while all other states are suppressed. Hence, in order to
implement  a quantum logic a strict control over both
moduli and phases of the amplitudes in superposition of states
in the register of a quantum computer is necessary. For
example, the quantum Control-Not (CN) gate CN$_{i,k}$ transforms
a superpositional wave function as
\begin{equation}
\label{supperposition1}
{\rm CN}_{i,k}^{\rm ideal}
\sum_{j=0}^{2^L-1}B_j|n_{L-1}\dots n_i\dots n_k\dots n_0\rangle=
\sum_{j=0}^{2^L-1}B_j
|n_{L-1}\dots n_i\dots n_i\oplus n_k\dots n_0\rangle,
\end{equation}
where $\oplus$ means sum modulo 2, $n_m=0,1$, and $L$ is the number of
qubits. The CN gate for remote qubits can be implemented using
a series of CN gates for the neighboring qubits, CN$_{i+1,i}$ or
CN$_{i,i+1}$.
(See Sec. \ref{sec:numerical} of this paper or,
for example, Ref.~\cite{perturbation}.)
Moreover, any quantum logic operation between remote qubits can be
divided into elementary logic gates between neighboring qubits. Hence,
in order to implement the universal quantum logic,
it is sufficient to construct only the universal quantum gates between
adjacent qubits.

Usually, in real quantum systems, instead of
Eq. (\ref{supperposition1}) one has
\begin{equation}
\label{supperposition2}
{\rm CN}_{i,k}^{\rm real}
\sum_{j=0}^{2^L-1}B_j|n_{L-1}\dots n_i\dots n_k\dots n_0\rangle=
\sum_{j=0}^{2^L-1}{B'}_j
|n_{L-1}\dots n_i\dots n_i\oplus n_k\dots n_0\rangle,
\end{equation}
where the coefficients ${B'}_j$ in the right-hand side of
Eq. (\ref{supperposition2}) are slightly different from the coefficients
$B_j$ in the right-hand side of Eq. (\ref{supperposition1}).
This introduces the error in the quantum algorithm.
A source of the error can be external, like noise, or
internal, like unwanted transitions driven by pulses of a protocol.

In this paper, we will consider only the latter errors.
We will present protocols for implementation of basic
quantum logic operations between the
neighboring qubits which work for an arbitrary superposition of
states in the Ising spin quantum computer. We use these operations
to implement the CN gate between the end qubits, CN$_{0,L-1}$ of the
spin chain, and estimate the phase and  probability errors.

\section{Ising spin quantum computer}
The simplest Hamiltonian for the Ising spin chain placed
in an external magnetic field can be represented as
\begin{equation}
\label{H}
H_n=-\sum_{k=0}^{L-1}\omega_kI_k^z
-2J\sum_{k=0}^{L-2}I_k^zI_{k+1}^z
-{\Omega_n\over 2}\sum_{k=0}^{L-1}
\left\{I_k^-\exp\left[-i\left(\nu_n t+\varphi_n\right)\right]+
h.c.\right\}
=H_0+V_n(t).
 \end{equation}
Here $\hbar=1$, $I_k^z$ is the operator
of the $z$ component of $k$th spin $1/2$, $I_k^\pm=I_k^x\pm I_k^y$,
$\omega_k$ is the Larmor frequency of the $k$th spin, $J$
is the interaction constant between the neighboring spins,
$\Omega_n$ is the Rabi frequency (frequency of precession around
the resonant transversal field in the rotating frame), $\nu_n$
is the frequency of the pulse, and
$\varphi_n$ is the phase constant of the $n$th pulse,
below called ``phase''.
The Hamiltonian (\ref{H}) is written for one $n$th
rectangular {\it rf} pulse. Below we omit the index $n$ which indicates
the pulse number.

In the interaction representation, the solution of the Schr\"odinger equation
can be written in the form
\begin{equation}
\label{Psi}
\Psi(t)=\sum_pC_p(t)|p\rangle\exp(-iE_pt),
\end{equation}
where $E_p$ and $|p\rangle$ are, respectively,
the eigenvalues and eigenfunctions of the Hamiltonian $H_0$.

Under the condition $\Omega<J\ll\omega$ the pulse effectively affects
only one $k$th spin in the chain~\cite{perturbation}
whose frequency $\omega_k$ is
close (near-resonant transition)
or equal (resonant transition) to the frequency of the pulse, $\nu$.
In this approximation, the system of coupled differential equations for
the coefficients $C_p(t)$ splits into $2^{L-1}$ independent groups.
Each group consists of two equations of the form
\begin{equation}
\label{dif_dynamics_phase}
~~i\dot C_p=-(\Omega/2)\exp[i(\Delta_{pm} t-\varphi)]C_m,
\end{equation}
$$
i\dot C_m=-(\Omega/2)\exp[-i(\Delta_{pm} t-\varphi)]C_p,
$$
where the states $|m\rangle$ and $|p\rangle$ are related by
the flip of $k$th spin,
$\Delta_{pm}=E_p-E_m-\nu$, and we suppose that $E_p>E_m$.
The solution of Eq. (\ref{dif_dynamics_phase}) is
%~\cite{9}
\begin{equation}
\label{2x2}
C_m(t_0+\tau)=\left[\cos(\lambda_{pm}\tau/2)+
i(\Delta_{pm}/\lambda)\sin(\lambda_{pm}\tau/2)\right]
\times\exp[-i\tau\Delta_{pm}/2],
\end{equation}
$$
C_p(t_0+\tau)=i(\Omega/\lambda_{pm})\sin(\lambda_{pm}\tau/2)\times
\exp[it_0\Delta_{pm}+i(\tau\Delta_{pm}/2-\varphi)].
$$
Here $t_0$ is the time of the beginning of the pulse,
$\lambda_{pm}=\sqrt{\Delta_{pm}^2+\Omega^2}$, $\tau$ is the
duration of the pulse, and the initial conditions are
$$
C_m(t_0)=1,\qquad C_p(t_0)=0.
$$
The solution for the transition from the upper state to the lower
state is
$$
C_m(t_0+\tau)=i(\Omega/\lambda_{pm})\sin(\lambda_{pm}\tau/2)\times
\exp[-it_0\Delta_{pm}-i(\tau\Delta_{pm}/2-\varphi)],
$$
\begin{equation}
\label{2x2a}
C_p(t_0+\tau)=\left[\cos(\lambda_{pm}\tau/2)-
i(\Delta_{pm}/\lambda)\sin(\lambda_{pm}\tau/2)\right]
\times\exp[i\tau\Delta_{pm}/2],
\end{equation}
$$
C_m(t_0)=0,\qquad C_p(t_0)=1.
$$

In the table below we present all states and their
energies for the spin chain of three qubits with the Larmor frequencies
$\omega_{0}=w$, $\omega_{1}=w+\delta\omega$,
$\omega_{2}=w+2\delta\omega$ ($w$ is the Larmor frequency of
the zeroth qubit) and with the Ising interaction
constant between the neighboring qubits~$J$,
\begin{equation}
\label{states}
\begin{array}{llll}
|0\rangle=|0_20_10_0\rangle,~&E_0=-\frac 32w-
\frac 32\delta\omega-J,\qquad\qquad &
|1\rangle=|001\rangle,~&E_1=-\frac 12w-\frac 32\delta\omega,  \\
|2\rangle=|010\rangle,&E_{2}=-\frac 12w-\frac 12\delta\omega+J,&
|3\rangle=|011\rangle,&E_{3}=\frac 12w-\frac 12\delta\omega,  \\
|4\rangle=|100\rangle,&E_{4}=-\frac 12w+\frac 12\delta\omega&
|5\rangle=|101\rangle,&E_{5}=\frac 12w+\frac 12\delta\omega+J,  \\
|6\rangle=|110\rangle,&E_{6}=\frac 12w+\frac 32\delta\omega&
|7\rangle=|111\rangle,&E_{7}=\frac 32w+\frac 32\delta\omega-J.  \\
\end{array}
\end{equation}
We will use this table for illustrative examples in the following
sections.

\section{The $2\pi k$-method}
Suppose, for example, that before the action of the pulse
we have the superposition of states $|0\rangle$ and $|4\rangle$
in Eq. (\ref{states}). In order to organize, for example, the transition
$|4\rangle\rightarrow|6\rangle$, associated with the flip of the
first (middle) spin, we apply the $\pi$ pulse
[$\Omega\tau=\pi$ in Eq. (\ref{2x2})] with the
resonant frequency $\nu=E_6-E_4=w+\delta\omega$. We also have
the near-resonant transition
(or transition with small detuning from the exact resonance condition)
$|0\rangle\rightarrow|2\rangle$
with the detuning $\Delta_{2,0}=E_2-E_0-\nu=2J$ which creates
the unwanted state with the probability [see Eq. (\ref{2x2})]
\begin{equation}
\label{P20}
P_{2,0}=\left[{\Omega\over\lambda_{2,0}}
\sin\left({\lambda_{2,0}\pi\over 2\Omega}\right)\right]^2,
\end{equation}
where
$\lambda_{2,0}=\sqrt{\Delta^2_{2,0}+\Omega^2}=\sqrt{4J^2+\Omega^2}$.
In order to suppress this transition we take the Rabi
frequency in the form
\begin{equation}
\label{Omegak}
\Omega_k={2J\over \sqrt{4k^2-1}},\qquad k=1,2,\dots.
\end{equation}
(See the $2\pi k$ method in Ref. \cite{9}.)
Then the argument of sinus in Eq. (\ref{P20}) turns to zero, and the
unwanted near-resonant transition $|0\rangle\rightarrow|2\rangle$
is completely suppressed.

\section{Generalized $2\pi k$ method}
Next, we will show that one cannot suppress all possible near-resonant
transitions using only one pulse. Since detunings
for different transitions are different, one $\pi$ pulse in general case
creates a relatively large error. In this section, we will
demonstrate how to compensate
this error using
additional correcting pulse. This procedure can be characterized
as a generalized $2\pi k$ method since this protocol suppresses
all near-resonant transitions for all quantum states in a
superposition.

Suppose that we have three quantum
states, $|0\rangle$, $|4\rangle$,
and $|5\rangle$ in a register of a quantum computer.
Using the resonant pulse we produce the resonant transition
$|0\rangle\rightarrow|2\rangle$ by flipping the first qubit.
We also have two
near-resonant transitions: $|4\rangle\rightarrow|6\rangle$ with the
detuning $\Delta_{6,4}=-2J$ and $|5\rangle\rightarrow|7\rangle$ with
$\Delta_{7,5}=-4J$. In order to suppress both transitions the Rabi
frequency $\Omega$ should satisfy the system of two equations
$$
\left|{\Delta_{6,4}\over\Omega}\right|=\sqrt{4k^2-1},\qquad
\left|{\Delta_{7,5}\over\Omega}\right|=\sqrt{4{K}^2-1},
$$
where $k$ and $K$ must be integer numbers. Dividing the second
equation over the first one and taking into consideration that
$|\Delta_{7,5}/\Delta_{6,4}|=2$ we obtain the relation
${K}^2=4k^2-3/4$. From this equation one can see that if $k$ is
integer then $K$ cannot be an integer number. Hence, if we
suppress, for example, the transition $|4\rangle\rightarrow|6\rangle$,
then we get an error for the transition
$|5\rangle\rightarrow|7\rangle$, and vice versa.

To proceed, we introduce the following notations.
The pulse $P_i^{00}$ indicates the $\pi$ pulse with the frequency
$\nu_i^{00}$ which is resonant for the transition
associated with the flip of $i$th spin when both its neighbors are
in the states $|0\rangle$. In a similar way, we define the pulses
$P_i^{10}$ and $P_i^{11}$.
The pulse $P_1^{00}$ will
be resonant, for example, for the transitions
$|0\rangle\rightarrow|2\rangle$ and $|2\rangle\rightarrow|0\rangle$
[see Eq. (\ref{states})].
We define $T_i^{00}$ as the transition (not necessarily resonant)
associated with the flip of $i$th spin when both its neighbors are in the
states $|0\rangle$. For example, the pulse $P_1^{00}$ gives rise
to the resonant
transitions $T_1^{00}$ and two types of the near-resonant transitions,
$T_1^{10}$ and $T_1^{11}$. In general case of an arbitrary number
of qubits, one notation, for example, $T_i^{00}$, means the transitions
for all quantum states with the identical groups
$|\dots0_{i+1}0_i0_{i-1}\dots\rangle$ and
$|\dots0_{i+1}1_i0_{i-1}\dots\rangle$. So,
now we discuss the types of transitions, but not the transitions between
definite states. We should note that
$\nu_i^{10}=\nu_i^{01}$, so that we use the same notation, $P_i^{10}$,
for the pulses $P_i^{10}$ and $P_i^{01}$.

Now we continue the discussion on the $2\pi k$ method.
All near-resonant transitions $T_i^{00}$ and $T_i^{11}$,
caused by the pulse $P_i^{10}$, can be
suppressed using only one pulse,
since the moduli of detunings are the same
and equal to $\Delta\equiv 2J$. In this case, we take
\begin{equation}
\label{deltaOmega}
{\Delta\over\Omega_1}=\sqrt{4k^2-1},
\end{equation}
where $k$ is the integer number.

As was shown above, pulses $P_i^{00}$ and
$P_i^{11}$ generate errors because the detunings
for transitions $T_i^{00}$, $T_i^{10}$ and $T_i^{11}$ are different.
To correct this error, we suppress the unwanted transition
$T_i^{11}$ for the pulse $P_i^{00}$ and
the unwanted transition
$T_i^{00}$ for the pulse $P_i^{11}$
with the moduli of detunings $\Delta_2\equiv2\Delta=4J$
by the $2\pi k$ method with
the Rabi frequency $\Omega_2$ satisfying the expression
\begin{equation}
\label{deltaOmega2}
\left|{\Delta_2\over\Omega_2}\right|=\sqrt{4k_2^2-1}.
\end{equation}
Below we take $k_2=k$ and $\Omega_2=2\Omega$, $\Omega_1=\Omega$,
where $\Omega$ is a parameter.
Thus, we choose the value of the Rabi frequency for the
pulses $P_i^{11}$ and $P_i^{00}$
twice larger than the value of the Rabi frequency for the
pulse $P_i^{10}$.

The pulses $P_i^{00}$ and $P_i^{11}$
with the parameters (\ref{deltaOmega2}) generate
errors in the result of the transitions $T_i^{10}$.
In Appendix \ref{sec:appendixA}, we demonstrate that
the additional pulse with the frequency $\nu_i^{10}$
can be used to correct this error.
The real and imaginary parts of the amplitude of the
created unwanted state are corrected by the proper choice
of the duration and the phase of the correcting pulse.
The correcting pulse does not generate errors since
the moduli of detunings for unwanted transitions,
$T_i^{00}$ and $T_i^{11}$, are the same, so that these transitions can
be suppressed simultaneously.

\section{Unwanted phases}
In spite of the fact that all pulses described above are
probability-corrected,
they generate unwanted phases which should also be compensated by
the protocol. In this section, we calculate all unwanted phases
which appear in the quantum computer in the result of action
of the {\it rf} pulses.

Even in the case when all unwanted near-resonant transitions
are suppressed, the unwanted phases appear because for definite
transitions $\Delta_{pm}\ne 0$ in Eqs. (\ref{2x2}) and (\ref{2x2a}).
The problem to compensate those phases becomes complicated,
since each pulse generates different phases for different states.
In this section we will show that
the unwanted phases can be compensated by choosing the proper
phases of the pulses of the protocol.
Each logic operation requires its own set of phases of the
pulses. However, if one has a set of phase-compensated universal gates,
introduced below, one can realize a quantum logic using these gates
as building blocks.

Further analysis is conducted in terms of the probability-corrected
pulses, $Q_i^{mn}(\varphi)$, where $m,n=0,1,$ $i=0,1,\dots L-1$, and
$\varphi$ is the phase of the pulse.
The pulse $Q_i^{10}(\varphi)$
indicates one $\pi$ pulse. It coincides with the pulse $P_i^{10}$,
introduced above.
The combined probability-corrected pulse,
$Q_i^{00}(\varphi)$ [or $Q_i^{11}(\varphi)$],
consists of one $\pi$ pulse $P_i^{00}$,
($P_i^{11}$), and one correcting pulse. The correcting
pulse does not implement quantum logic, it only removes unwanted states
from the register of quantum computer.

In the Table I we present the phases (see Appendix \ref{sec:appendixB})
acquired by different
states (which can be in a superposition in a register of
quantum computer) generated in the result of the different kinds
of the probability-corrected pulses, $Q_i^{mn}(\varphi)$.

\begin{table}[hbt]
\begin{center}
\begin{tabular}{|c|c|c|c|c} \cline{1-4}
 state   &  \multicolumn{3}{|c|}{acquired phase} \\ \cline{2-4}
 &   $Q_i^{01}(\varphi)$  &$Q_i^{00}(\varphi)$&$Q_i^{11}(\varphi)$&
 \\ \cline{1-4}
$|\dots 0_{i+1}0_i0_{i-1}\dots\rangle$&   $-\theta$&
$\pi/2-\varphi+\gamma^*$&   $-\theta-\gamma$ \\

$|\dots 0_{i+1}1_i0_{i-1}\dots\rangle$&   $\theta$&
$\pi/2+\varphi-\gamma^*$&   $\theta+\gamma$ \\

$|\dots 1_{i+1}0_i0_{i-1}\dots\rangle$&   $\pi/2-\varphi^*$&
$\pi+\theta/2+\Theta$&   $\pi-\theta/2-\Theta$ \\

$|\dots 1_{i+1}1_i0_{i-1}\dots\rangle$&   $\pi/2+\varphi^*$&
$\pi-\theta/2-\Theta$&  $\pi+\theta/2+\Theta$ \\

$|\dots 0_{i+1}0_i1_{i-1}\dots\rangle$&   $\pi/2-\varphi^*$&
$\pi+\theta/2+\Theta$&   $\pi-\theta/2-\Theta$ \\

$|\dots 0_{i+1}1_i1_{i-1}\dots\rangle$&   $\pi/2+\varphi^*$&
$\pi-\theta/2-\Theta$&   $\pi+\theta/2+\Theta$ \\

$|\dots 1_{i+1}0_i1_{i-1}\dots\rangle$&   $\theta$&
$\theta+\gamma$&   $\pi/2-\varphi-\gamma^*$ \\

$|\dots 1_{i+1}1_i1_{i-1}\dots\rangle$&   $-\theta$&
$-\theta-\gamma$&   $\pi/2+\varphi+\gamma^*$ \\
\cline{1-4}
\end{tabular}

\caption{Phases generated by the probability-corrected
pulses, $Q_i^{mn}(\varphi)$.
The asterisk indicates that the resonant transition from the state
shown in the first column of the table to the other
state, associated with the flip of the $i$th qubit, takes place.
The phases, $\theta$, $\Theta$ and $\gamma$, are defined
in the Appendices \ref{sec:appendixA} and \ref{sec:appendixB}.
}\label{tab:table1}
\end{center}
\end{table}

\section{Phase and probability-corrected universal gates}
In the Ising spin quantum computer the parameters which allow one to
compensate the unwanted phases are the phases of the
{\it rf} pulses. From Eq. (\ref{2x2}) one can see that the
phase $\varphi$ of the pulse changes the
phase of the wave function. In this section we provide the phase
and probability correct protocols for realization of universal
quantum gates on arbitrary superposition of states
in the Ising spin scalable quantum computer.

Our task is to construct phase-corrected quantum gates from
the elementary probability-corrected three-qubit elementary gates,
$Q^{mn}_i(\varphi)$, $m,n=0,1$, $i=0,1,\dots,L-1$. These gates act on the
target qubit $q_i$, for given configurations of its nearest
neighbors, $q_{i-1}$ and $q_{i+1}$.
We thus define the operations
$Q^{mn}_i(\varphi)$ in terms of their effects on the states,
$|\dots q_{i+1}q_iq_{i-1}\dots\rangle$ as
\begin{itemize}
  \item for $Q^{01}_i(\varphi)\equiv Q^{10}_i(\varphi)$,
  {\setlength\arraycolsep{2pt}\begin{eqnarray}
    &\ &|\dots 0_{i+1}q_i0_{i-1}\dots\rangle\rightarrow\exp
    i[(-1)^{q_i}\,(-\theta)]|0_{i+1}q_i0_{i-1}\dots\rangle,
    \nonumber \\
    &\ &|\dots 0_{i+1}q_i1_{i-1}\dots\rangle\rightarrow\exp
    i\left[\pi/2+(-1)^{q_i}\,(-\varphi)
    \right]|\dots 0_{i+1}\overline{q_i}1_{i-1}\dots\rangle,
    \nonumber \\
    &\ &|\dots 1_{i+1}q_i0_{i-1}\dots\rangle\rightarrow\exp
    i\left[\pi/2+(-1)^{q_i}\,(-\varphi)
    \right]|\dots 1_{i+1}\overline{q_i}0_{i-1}\dots\rangle,
    \nonumber \\
    &\ &|\dots 1_{i+1}q_i1_{i-1}\dots\rangle\rightarrow\exp
    i[(-1)^{q_i}\,(\theta)]|\dots1_{i+1}q_i1_{i-1}\dots\rangle,
  \end{eqnarray}}
  \item for $Q^{00}_i(\varphi)$,
  {\setlength\arraycolsep{2pt}\begin{eqnarray}
    &\ &|\dots0_{i+1}q_i0_{i-1}\dots\rangle\rightarrow\exp
    i\left[\pi/2+(-1)^{q_i}(-\varphi+\gamma)
    \right]|\dots0_{i+1}\overline{q_i}0_{i-1}\dots\rangle,
    \nonumber \\
    &\ &|\dots0_{i+1}q_i1_{i-1}\dots\rangle\rightarrow\exp
    i\left[\pi+(-1)^{q_i}\,(\theta/2+\Theta)\right]
    |\dots0_{i+1}q_i1_{i-1}\rangle,
    \nonumber \\
    &\ &|\dots1_{i+1}q_i0_{i-1}\dots\rangle\rightarrow\exp
    i\left[\pi+(-1)^{q_i}\,(\theta/2+\Theta)\right]
    |\dots1_{i+1}q_i0_{i-1}\dots\rangle,
    \nonumber \\
    &\ &|\dots 1_{i+1}q_i1_{i-1}\dots\rangle\rightarrow\exp
    i\left[(-1)^{q_i}\,(\theta+\gamma)\right]
    |\dots0_{i+1}q_i0_{i-1}\dots\rangle,
  \end{eqnarray}}
  \item for $Q^{11}_i(\varphi)$,
  {\setlength\arraycolsep{2pt}\begin{eqnarray}
    &\ &|\dots0_{i+1}q_i0_{i-1}\dots\rangle\rightarrow\exp
    i\left[(-1)^{q_i}\,(-\theta-\gamma)\right]
    |\dots0_{i+1}q_i0_{i-1}\dots\rangle,
    \nonumber \\
    &\ &|\dots0_{i+1}q_i1_{i-1}\dots\rangle\rightarrow\exp
    i\left[\pi+(-1)^{q_i}\,(-\theta/2-\Theta)\right]
    |\dots0_{i+1}q_i1_{i-1}\dots\rangle,
    \nonumber \\
    &\ &|\dots1_{i+1}q_i0_{i-1}\dots\rangle\rightarrow\exp
    i\left[\pi+(-1)^{q_i}\,(-\theta/2-\Theta)\right]
    |\dots1_{i+1}q_i0_{i-1}\dots\rangle,
    \nonumber \\
    &\ &|\dots1_{i+1}q_i1_{i-1}\dots\rangle\rightarrow\exp
    i\left[\pi/2+(-1)^{q_i}(-\varphi-
    \gamma)\right]|\dots1_{i+1}\overline{q_i}1_{i-1}\dots\rangle.
  \end{eqnarray}}
\end{itemize}
One can see that each $Q_i^{mn}$ pulse introduces three
different kinds of phases into different states.  These phases arise from
different types of resonant and near-resonant transitions
initiated by the pulses.

\subsection{Not gate}
Each pulse has one externally
controllable phase, $\varphi$, which can influence
different subsets of states.  Using combinations of 
three pulses on qubit $i$, one can introduce at most three
independent phases to correct different unwanted
phases generated for different states in the register of
a quantum computer.

For the Not gate, the correct transformation for the amplitudes of
the states can be implemented by the sequence $Q_i^{11}(\varphi_3)
Q_i^{01}(\varphi_2)Q_i^{00}(\varphi_1)$. (The order of
implementation of the operators is from the right to the left.) The
transformation which results from this sequence of operations is,
{\setlength\arraycolsep{2pt}\begin{eqnarray}
  &\ &|\dots 0_{i+1}q_i0_{i-1}\dots\rangle\rightarrow\exp
  i\left[\pi/2+(-1)^{q_i}\left(-\varphi_1+2\gamma+2\theta\right)\right]
  |\dots0_{i+1}\overline{q_i}0_{i-1}\dots\rangle,
  \nonumber \\
  &\ &|\dots0_{i+1}q_i1_{i-1}\dots\rangle\rightarrow\exp
  i\left[\pi/2+(-1)^{q_i}(-\varphi_2+\theta+2\Theta)
  \right]|\dots0_{i+1}\overline{q_i}1_{i-1}\dots\rangle,
  \nonumber \\
  &\ &|\dots1_{i+1}q_i0_{i-1}\dots\rangle\rightarrow\exp
  i\left[\pi/2+(-1)^{q_i}(-\varphi_2+\theta+2\Theta)
  \right]|\dots1_{i+1}\overline{q_i}0_{i-1}\dots\rangle,
  \nonumber \\
  &\ &|\dots1_{i+1}q_i1_{i-1}\dots\rangle\rightarrow\exp
  i\left[\pi/2+(-1)^{q_i}\left(-\varphi_3+2\theta
  \right)\right]|\dots1_{i+1}\overline{q_i}1_{i-1}\dots\rangle.
\end{eqnarray}}
It is easy to see, that the implementation of the Not gate
requires the following set of phases:
$\varphi_1=2\gamma+2\theta$,
$\varphi_2=\theta+2\Theta$ and $\varphi_3=2\theta$.
The overall phase factor is equal to $\pi/2$.

\subsection{Control-Not gate}
For qubits with homogeneous
coupling between them, the correct CN$_{a,b}$ gate, where $a$ is the
number of the control qubit and $b=a\pm 1$ is the number of the target
qubit, can be implemented by the following sequence,
{\setlength\arraycolsep{2pt}\begin{eqnarray}\label{eq:CN21}
{\rm CN}_{a,b}&=&Q_a^{11}(0)Q_a^{10}(0)
Q_a^{00}(\varphi_8)Q_{b}^{01}(0)
Q_{b}^{01}(\varphi_7)
Q_{b}^{00}(\varphi_6) \nonumber \\
& \ & Q_a^{11}(\varphi_5)Q_a^{10}(\varphi_4)
Q_a^{00}(\varphi_3)Q_{b}^{01}(0)Q_{b}^{01}(\varphi_2)
Q_{b}^{11}(\varphi_1),
\end{eqnarray}}
where the phases $\varphi_n$, are computed in
Appendix {\ref{sec:AppendixC},

\begin{tabular}{llll}
$\varphi_1=-5\theta-2\gamma$,~~~&
 $\varphi_2=\frac 52\theta-\Theta+\gamma$,~~~&
 $\varphi_3=\frac{3}{4}\pi+2\theta-4\Theta+2\gamma$,~~~&
 $\varphi_4=\frac{3}{4}\pi$,\\
 $\varphi_5=\frac{3}{4}\pi$,&
 $\varphi_6=-2\Theta$,&
 $\varphi_7=-\frac 52\theta+\Theta-\gamma$,&
 $\varphi_{8}=2\theta-4\Theta+2\gamma$.
\end{tabular}

\vspace{3mm}
In order to illustrate the action of the CN gate, in the Table II
we show how each pulse of the gate CN$_{i+1,i}$
modifies the phase (initially equal to zero) of the state
$|\dots 0_{i+2}1_{i+1}0_i0_{i-1}\dots\rangle$ due to the Table I.
In a similar way, one can show that other states
acquire the same phase $\pi/4$ after the action of the CN gate,
while their phases in the middle of the CN gate protocol
can be different.

\begin{table}
\begin{tabular}{|c|c|c|c|c|}
%$$
%\begin{array}{|c|c|c|c|c|}
\hline
& pulse & produced state & acquired phase & phase\\
$1$&$Q_i^{11}\left(-5\theta-2\gamma\right)$&
$|\dots 0_{i+2}1_{i+1}0_i0_{i-1}\dots\rangle$&
$\pi-\frac 12 \theta-\Theta$& $\pi-\frac 12 \theta-\Theta$\\
$2$&$Q_i^{10}\left(\frac 52\theta-\Theta+\gamma\right)$&
$|\dots 0_{i+2}1_{i+1}1_i0_{i-1}\dots\rangle$&
$\frac 12\pi-\left(\frac 52\theta-\Theta+\gamma\right)$&
$\frac 32\pi-3\theta-\gamma$\\
$3$&$Q_i^{10}\left(0\right)$&
$|\dots 0_{i+2}1_{i+1}0_i0_{i-1}\dots\rangle$&
$\frac 12\pi$& $-3\theta-\gamma$\\
$4$&$Q_{i+1}^{00}\left(\frac 34\pi+2\theta-\right.$&
$|\dots 0_{i+2}0_{i+1}0_i0_{i-1}\dots\rangle$&
$\frac 12\pi+\frac 34\pi+2\theta-$&
$\frac 54\pi-\theta-4\Theta$\\
&$\left.\qquad 4\Theta+2\gamma\right)$& & $\qquad 4\Theta+2\gamma$&\\
$5$&$Q_{i+1}^{10}\left(\frac 34\pi\right)$&
$|\dots 0_{i+2}0_{i+1}0_i0_{i-1}\dots\rangle$&
$-\theta$&$ \frac 54\pi-2\theta-4\Theta$\\
$6$&$Q_{i+1}^{11}\left(\frac 34\pi\right)$&
$|\dots 0_{i+2}0_{i+1}0_i0_{i-1}\dots\rangle$&
$-\theta-\gamma$&
$\frac 54\pi-3\theta-4\Theta-\gamma$\\
$7$&$Q_{i}^{00}\left(-2\Theta\right)$&
$|\dots 0_{i+2}0_{i+1}1_i0_{i-1}\dots\rangle$&
$\frac 12\pi+2\Theta+\gamma$&
$\frac 74\pi-3\theta-2\Theta$\\
$8$&$Q_{i}^{10}\left(-{5\over 2}\theta+\Theta-\gamma\right)$&
$|\dots 0_{i+2}0_{i+1}1_i0_{i-1}\dots\rangle$&
$\theta$&
$\frac 74\pi-2\theta-2\Theta$\\
$9$&$Q_{i}^{10}\left(0\right)$&
$|\dots 0_{i+2}0_{i+1}1_i0_{i-1}\dots\rangle$&
$\theta$&
${7\over 4}\pi-\theta-2\Theta$\\
$10$&$Q_{i+1}^{00}\left(2\theta-4\Theta+2\gamma\right)$&
$|\dots 0_{i+2}0_{i+1}1_i0_{i-1}\dots\rangle$&
$\pi+\frac 12\theta+\Theta$&
${3\over 4}\pi-\frac 12\theta-\Theta$\\
$11$&$Q_{i+1}^{10}\left(0\right)$&
$|\dots 0_{i+2}1_{i+1}1_i0_{i-1}\dots\rangle$&
${1\over 2}\pi$&
${5\over 4}\pi-\frac 12\theta-\Theta$\\
$12$&$Q_{i+1}^{11}\left(0\right)$&
$|\dots 0_{i+2}1_{i+1}1_i0_{i-1}\dots\rangle$&
$\pi+\frac 12\theta+\Theta$&
${1\over 4}\pi$\\
\hline
\end{tabular}
%$$
\caption{\label{tab:table4}Modification of the phase
of the initial state $|\dots 0_{i+2}1_{i+1}0_i0_{i-1}\dots\rangle$
in the result of the action of the CN$_{i+1,i}$ gate pulses.}
\end{table}

\subsection{Not and Control-Not gate on edge qubits}
\vspace{-3mm}
In order to make our set of elementary quantum logic operations
complete, we also present the protocols for the logic gates
on the edge qubits. The Not gate is implemented by
the sequence of two pulses:
\vspace{-2mm}
\begin{equation}
\label{Not}
{\rm Not}_i=Q_i^1(\theta)Q_i^0(\theta),
\end{equation}
where $i=0,L-1$.
In Eq. (\ref{Not}) we suppose that the notation $Q_i^m(\varphi)$,
$m=0,1$, with one upper index
means one $\pi$ pulse with the corresponding
resonant frequency and the phase $\varphi$. The overall phase
for the gate (\ref{Not}) is $\pi/2$. The gate CN$_{a,b}$
with the edge target qubit $b$ $(b=0~{\rm or}~L-1)$ is
\vspace{-2mm}
\begin{equation}
\label{CNot}
{\rm CN}_{a,b}=Q_a^{11}(0)Q_a^{10}(0)Q_a^{00}(0)
Q_a^{11}\left({1\over 4}\pi\right)Q_a^{10}\left({1\over 4}\pi\right)
Q_a^{00}\left({1\over 4}\pi\right)Q_b^{0}(0)Q_b^{0}(-\theta)
Q_b^{1}(-2\theta),
\end{equation}
with the overall phase $-{\pi\over 4}$.
The gate CN$_{a,b}$,
with the edge control qubit $a$ $(a=0~{\rm or}~L-1)$, is
\begin{equation}
\label{CNot1}
{\rm CN}_{a,b}=Q_a^{1}(0)Q_a^{0}(0)Q_b^{10}(0)Q_b^{10}(0)
Q_b^{00}\left(-6\theta+2\Theta-2\gamma\right)
Q_a^{1}\left({3\over 4}\pi+{5\over 2}\theta-\Theta+\gamma\right)
\times
\end{equation}
$$
Q_a^{0}\left({3\over 4}\pi-{5\over 2}\theta+\Theta-\gamma\right)
Q_b^{10}(0)Q_b^{10}(5\theta-2\Theta+2\gamma)
Q_b^{11}\left(-2\Theta\right),
$$
with the overall phase $\pi\over 4$.

\subsection{A single qubit rotations around $x$ axis
for an arbitrary angle}
We denote the pulses for rotation of $i$th spin
around $x$ axis for the angle
$\rho\pi/2$ as $Q_{i\rho}^{mn}(\varphi)$, $m,n=0,1$, $i=1,\dots,L-2$ for
intermediate qubits, and $Q_{i\rho}^{m}(\varphi)$, $i=0,L-1$ for
the edge qubits. Note that the case $\rho=1$
was analyzed before.
From the $2\pi k$-condition for $\Omega_\rho$ one obtains
\begin{align}
\theta_\rho&=\pi\sqrt{k^2-\rho^2/4},\qquad
&\alpha_\rho&=\frac{\pi}{2}\sqrt{k^2+3\rho^2/4},\\
\tan\Theta_\rho &=-2\frac{\theta_\rho}{\alpha_\rho}\tan\alpha_\rho,
&\tan \beta_\rho  &= -\frac{\pi}{2 \alpha_\rho}
\tan \alpha_\rho \cos \Theta_\rho
\end{align}
and
\begin{equation}
%\beta_\rho^*  &=\beta_\rho+\pi,\quad
\gamma_\rho = \sqrt{(\pi k)^2-(\pi +\beta_\rho )^2},
\end{equation}
where we took the same integer value, $k_\rho=k$, as before.
The pulse $Q_{i\rho}^{10}(\varphi)$
is the single pulse with the Rabi frequency
$\Omega_{1,\rho}=\Delta\rho/\sqrt{4 k^2-\rho^2}\equiv\Omega_{\rho}$,
duration $\tau_\rho=\rho\pi/\Omega_{\rho}$,
frequency $\nu_i^{10}$, and phase $\varphi$.
Note, that the Rabi frequency, $\Omega_{\rho}$, decreases with
$\rho$ decreasing. The pulse
$Q_{i\rho}^{11}(\varphi)$ [or
$Q_{i\rho}^{00}(\varphi)$] is composed of the pulse with
the Rabi frequency $\Omega_{2,\rho}=2 \Omega_{\rho}$,
duration $\tau=\rho\pi/\Omega_{2,\rho}$,
frequency $\nu_{i}^{11}$ ($\nu_i^{00}$), and the phase  $\varphi$,
and a correcting pulse with the Rabi frequency
$\Omega_{{c},\rho}=\Delta(\beta_\rho+\pi)/\gamma_\rho $,  duration
$\tau_{{c},\rho}=2\gamma_\rho/\Delta$, frequency
$\nu_i^{10}$, and the phase
$\varphi^{11}_{{c},\rho}=
-\theta_\rho+ \varphi - \Delta t_0 -\Theta_\rho$
($\varphi^{00}_{{c},\rho}=\theta_\rho+ \varphi + \Delta t_0 +\Theta_\rho$).
The phases generated by the pulses
for $Q_{i\rho}^{mn}(\varphi)$ are the same as
the ones indicated in the Table I, if we replace
$\theta$ with $\theta_\rho$, $\Theta$ with
$\Theta_\rho$, and $\gamma$ with $\gamma_\rho$.
The resonant transitions for an arbitrary angle
create a superposition of states. In the Table \ref{tab:fases} we
specify the phases of both, excited and initial states after the transition.
\begin{table}[h]
\begin{center}
\begin{tabular}{|c|c|c|c|}
\hline
Pulse & Initial state & Phase of initial state & Phase of excited state \\ \hline
$Q_{i\rho}^{00}(\varphi)$ & $|\dots 0_{i+1}0_i0_{i-1} \dots \rangle$
& $-\gamma_\rho$ & $\pi/2-\varphi+\gamma_\rho$ \\
$Q_{i\rho}^{00}(\varphi)$ & $|\dots 0_{i+1}1_i0_{i-1} \dots \rangle$
& $\pi/2+\varphi-\gamma_\rho$ & $\gamma_\rho$ \\
$Q_{i\rho}^{10}(\varphi)$ & $|\dots 0_{i+1}0_i1_{i-1} \dots \rangle$
& $0$ & $\pi/2- \varphi$ \\
$Q_{i\rho}^{10}(\varphi)$ & $|\dots 0_{i+1}1_i1_{i-1} \dots \rangle$
& $\pi/2 + \varphi$ & $0$ \\
$Q_{i\rho}^{10}(\varphi)$ & $|\dots 1_{i+1}0_i0_{i-1} \dots \rangle$
& $0$ & $\pi/2- \varphi$ \\
$Q_{i\rho}^{10}(\varphi)$ & $|\dots 1_{i+1}0_i1_{i-1} \dots \rangle$
& $\pi/2 + \varphi$ & $0$ \\
$Q_{i\rho}^{11}(\varphi)$ & $|\dots 1_{i+1}0_i1_{i-1} \dots \rangle$
& $\gamma_\rho$ & $\pi/2-\varphi-\gamma_\rho$ \\
$Q_{i\rho}^{11}(\varphi)$ & $|\dots 1_{i+1}1_i1_{i-1} \dots \rangle$
& $\pi/2+\varphi+\gamma_\rho$ & $-\gamma_\rho$ \\
\hline
\end{tabular}
\end{center}
\caption{Phases acquired after a $Q_{i\rho}^{nm}$ pulse for resonant transitions.}
\label{tab:fases}
\end{table}

The probability and phase correct rotation of a single qubit can
be implemented by the operators:
%\cite{libroberman}:
\begin{equation}
  \begin{split}
    U_j^\rho(\varphi)|\dots 0_j\dots\rangle&=\cos(\rho\pi/2)
    |\dots 0_j\dots\rangle+
    ie^{i\varphi}\sin(\rho\pi/2)|\dots 1_j\dots\rangle,\\
    U_j^\rho(\varphi)|\dots 1_j\dots\rangle&=
    \cos(\rho\pi/2)|\dots 1_j\dots\rangle+
    ie^{-i\varphi}\sin(\rho\pi/2)|\dots 0_j\dots\rangle.
  \end{split}
\end{equation}
The operators $U_j^\rho(\varphi)$ can be build
using the following sequence of pulses $Q_{i\rho}^{mn}(\varphi)$:

\noindent
for intermediate qubits:
\begin{multline}
U_j^\rho(\varphi)= Q_{j\rho}^{10}(\varphi)
      Q_{j}^{00}(2(\gamma +2\theta + \varphi +\gamma_\rho+\theta_\rho))
      Q_{j}^{00}(0)
      Q_{j\rho}^{00}(-4\gamma- 8\theta + \varphi - 2\gamma_\rho- 2\theta_\rho)\\
      \times
      Q_{j}^{11}(-2(\gamma +2\theta + \gamma_\rho+\theta_\rho))
      Q_{j}^{11}(0)
      Q_{j\rho}^{11}(4\theta+\varphi)
      Q_{j}^{10}(0)
      Q_{j}^{10}(0);
\end{multline}

\noindent
for the edge qubits:

\begin{equation}
U_j^\rho(\varphi)= Q_{j\rho}^{1}(\varphi) Q_{j\rho}^{0}(\varphi+2\theta_\rho)
      Q_{j}^{1}(-\theta_\rho) Q_{j}^{0}(\theta_\rho) Q_{j}^{1}(0) Q_{j}^{0}(0),
\end{equation}
with the overall phase $\pi$.

\section{Numerical results}
\label{sec:numerical}
In the above sections we were concerned only with the errors
generated in the result of the near-resonant transitions
with frequencies close to the frequency of the external
field, when $E_p-E_m-\nu\sim J$. These transitions are
associated with the flip of the resonant $k$th spin whose the Larmor
frequency, $\omega_k$, is close
to the frequency of the external field, $\omega_k\approx \nu$.
In general case, the near-resonant transitions generate
large errors. In the preceding sections the procedure is
described how to suppress all these transitions.
However, since the external {\it rf} field affects all spins
in the system, there are non-resonant transitions associated with
flips of other non-resonant $k'$th spins.
Because the non-resonant transitions
$|p'\rangle\rightarrow|m'\rangle$
have large detuning, $E_{p'}-E_{m'}-\nu\sim \delta\omega|k-k'|$,
their probabilities are small and proportional to $\mu^2/|k-k'|^2$,
where~\cite{perturbation,BI3}
\begin{equation}
\label{mu}
\mu=(\Omega/2\delta\omega).
\end{equation}
For typical parameters $\mu\sim 10^{-4}$. Since in the system
there are only non-resonant transitions
(and all near-resonant transitions are suppressed),
for a definite protocol $\mu$ is the only parameter which defines
the errors in our quantum computer. This is demonstrated in the numerical
simulations below.

\begin{figure}
\mbox{
\includegraphics[width=8cm,height=8cm]{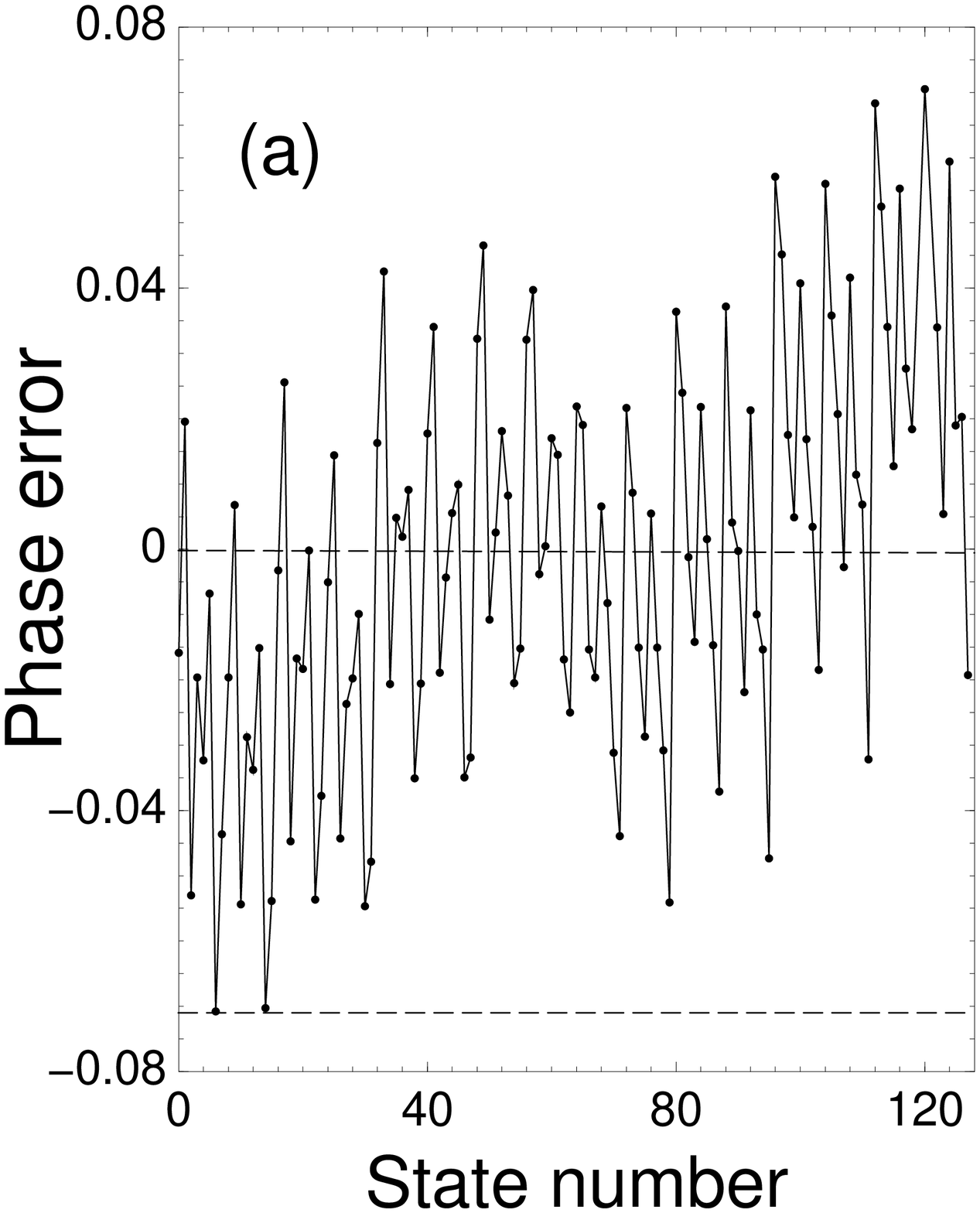}
\includegraphics[width=8cm,height=8cm]{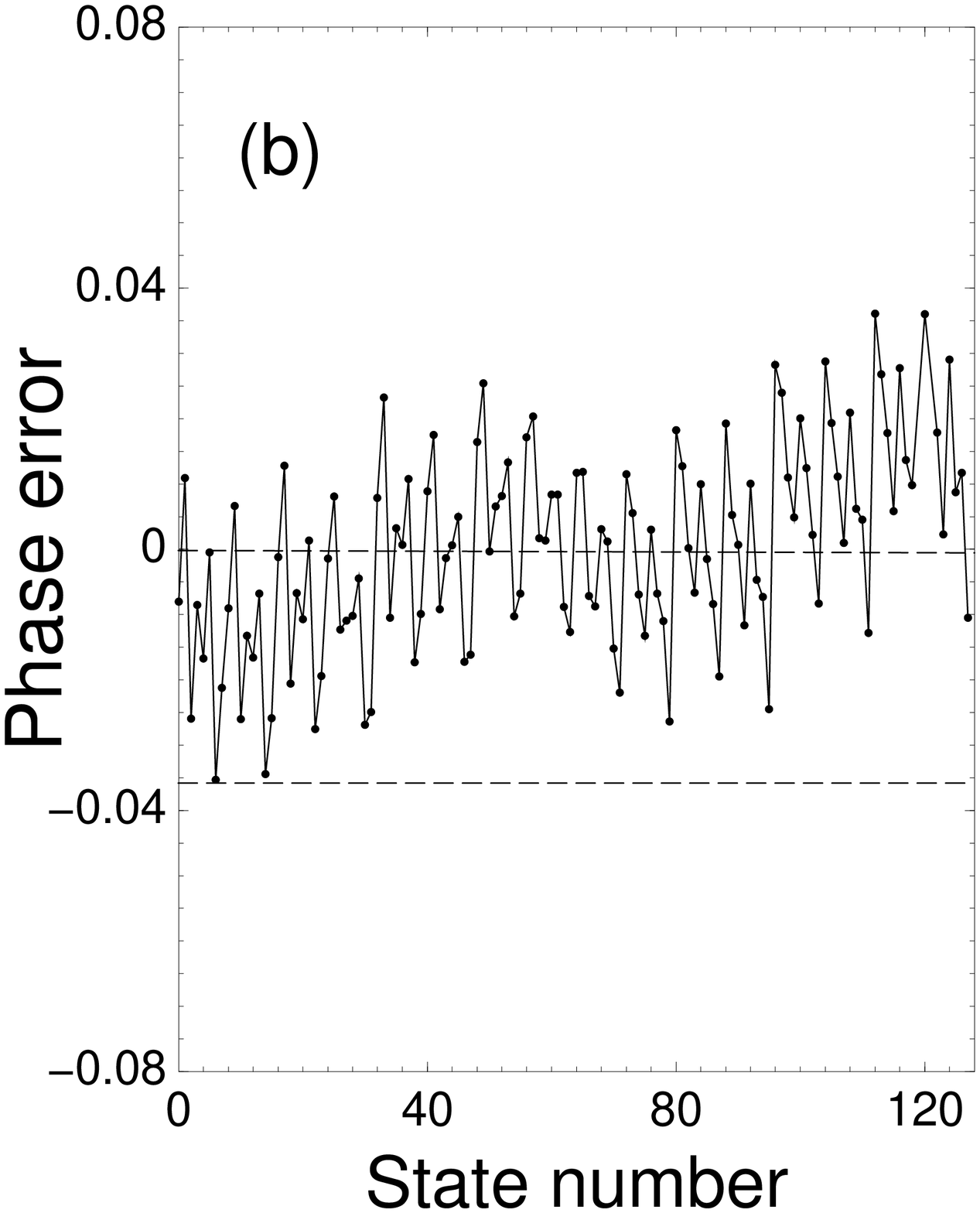}}
\vspace{-5mm}
\caption{Deviation of phases from the common phase
for different states of the superposition after implementation
of the CN$_{0,L-1}$ gate, $L=7$, $\Omega=\Omega_k$
[see Eq. (\ref{Omegak})], $k=2$, $J=1$;
(a) $\delta\omega=10^4$, (b) $\delta\omega=2\times 10^4$.}
\label{fig:1}
\end{figure}

In order to demonstrate the action of the logic gates on superpositional
states, described in this paper, we simulated the exact quantum
dynamics~\cite{perturbation}
of the system with the Hamiltonian (\ref{H})
during implementation of the CN
gate (\ref{supperposition1}) between the edge qubits of the spin chain,
CN$_{0,L-1}$, where the $0$th qubit is the control qubit and the $(L-1)$th qubit
is the target qubit. (In a similar way one can implement
any other CN gate CN$_{i,j}$, where
$i,j=0,\dots,L-1$ and $i\ne j$, for example CN$_{L-1,0}$.)
Using the Swap operations,
\begin{equation}
\label{swap}
{\rm S}_{i,i+1}={\rm S}_{i+1,i}=
{\rm CN}_{i,i+1}{\rm CN}_{i+1,i}{\rm CN}_{i,i+1}=
{\rm CN}_{i+1,i}{\rm CN}_{i,i+1}{\rm CN}_{i+1,i},
\end{equation}
we move the control $0$th qubit to the $L-2$th position,
implement the CN gate
${\rm CN}_{L-2,L-1}$, and using the same Swap gates we return the
control qubit to its initial $0$th position. The whole procedure
can be written in the form (read from the right to the left)
\begin{equation}
\label{cn0L}
{\rm CN}_{0,L-1}={\rm S}_{0,1}{\rm S}_{1,2}
\dots{\rm S}_{L-4,L-3}{\rm S}_{L-3,L-2}{\rm CN}_{L-2,L-1}
{\rm S}_{L-2,L-3}{\rm S}_{L-3,L-4}
\dots{\rm S}_{2,1}{\rm S}_{1,0}.
\end{equation}

\begin{figure}
\includegraphics[width=10cm,height=8cm]{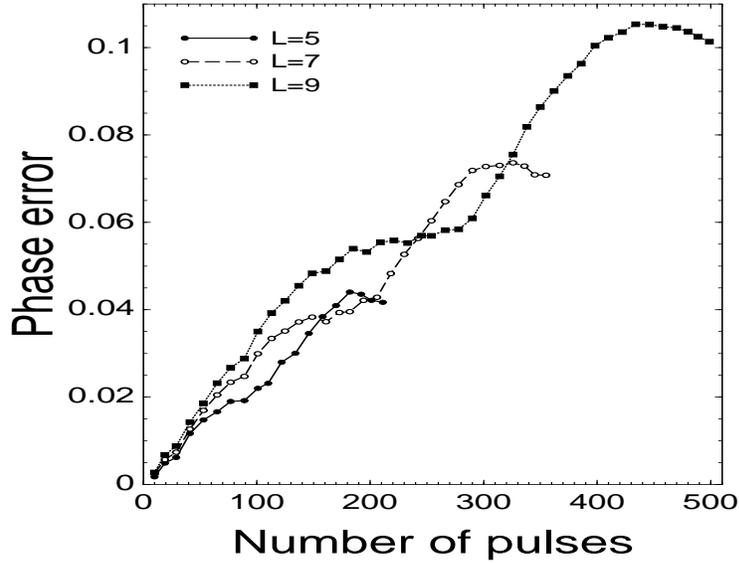}
\vspace{-5mm}
\caption{The phase error as a function of the number of pulses
for four values of $L$. $\delta\omega=10^4$, other parameters
are the same as in Figs. 1(a,b).}
\label{fig:2}
\end{figure}

For simulations we initialized the system in the superpositional state
(\ref{supperposition1}) with randomly chosen real positive values of the
coefficients $B_j$, subject only to the normalization condition
$\sum_{j=0}^{2^L-1}B_j^2=1$. Since all $B_j$ are real and positive
the phases
of all states of the superposition are the same and equal to zero.
(We should note that our protocols work for any
set of initial phases.)

In Figs. 1(a,b) we show the deviation of phases
for different states,
$\varphi_j-\Phi$, $j=0,\dots,2^L-1$ [here $\Phi$ is the total
phase and $\varphi_j={\rm arctan}[{\rm Im}({B'}_j)/{\rm Re}({B'}_j))]$,
where Im and Re stand for the imaginary and real
parts of the amplitude ${B'}_j$ in Eq. (\ref{supperposition2})],
from the total phase $\Phi$
after implementation of the CN$_{0,L-1}$ gate
for two values of $\delta\omega$
[and correspondent values of $\mu$ in Eq. (\ref{mu})].
From comparison of Fig. 1(a) with Fig. 1(b) one can see that
decreasing twice the value of $\mu$ leads to decrease of
the error in the phases
of all states by the factor two. This supports the fact, mentioned
in the
beginning of this section, that for our protocols
$\mu$ is the only parameter which defines the errors in
the quantum computer. Physically, this means that only the non-resonant
transitions contribute to the errors and all near-resonant
transitions are completely suppressed.

\begin{figure}
\mbox{\includegraphics[width=8cm,height=6cm]{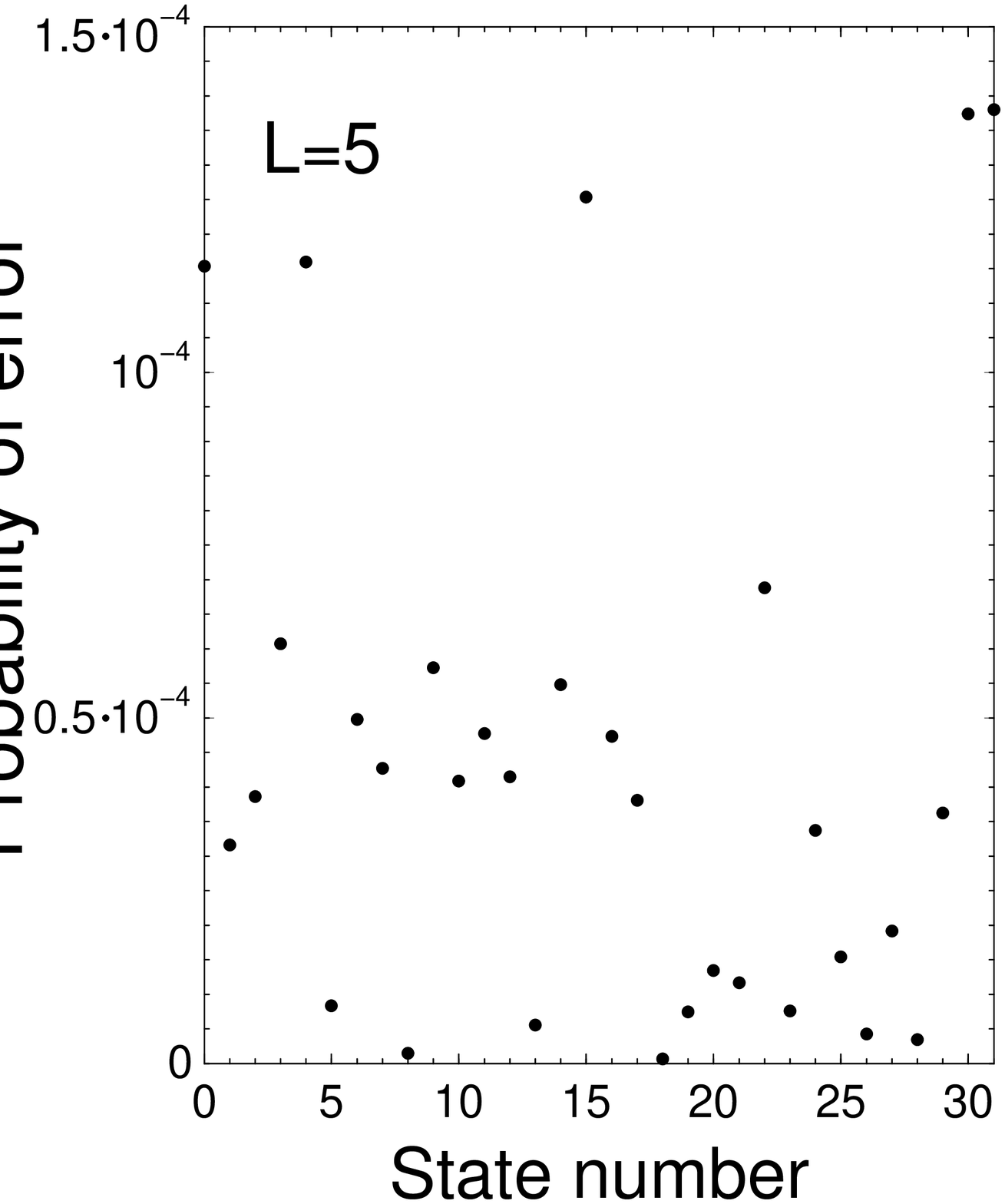}~
\includegraphics[width=8cm,height=6cm]{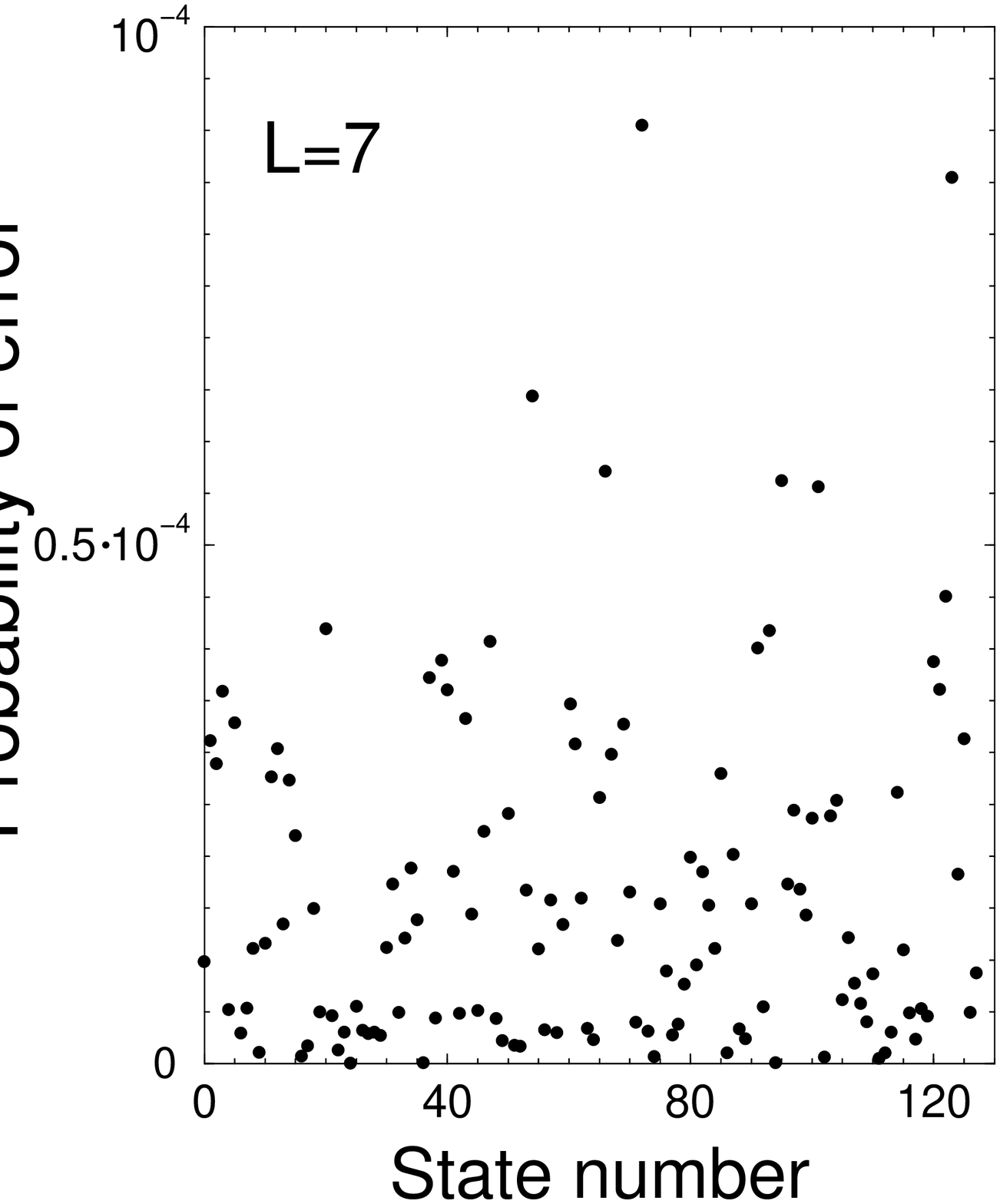}}
%\vspace{-10mm}
\mbox{\includegraphics[width=8cm,height=6cm]{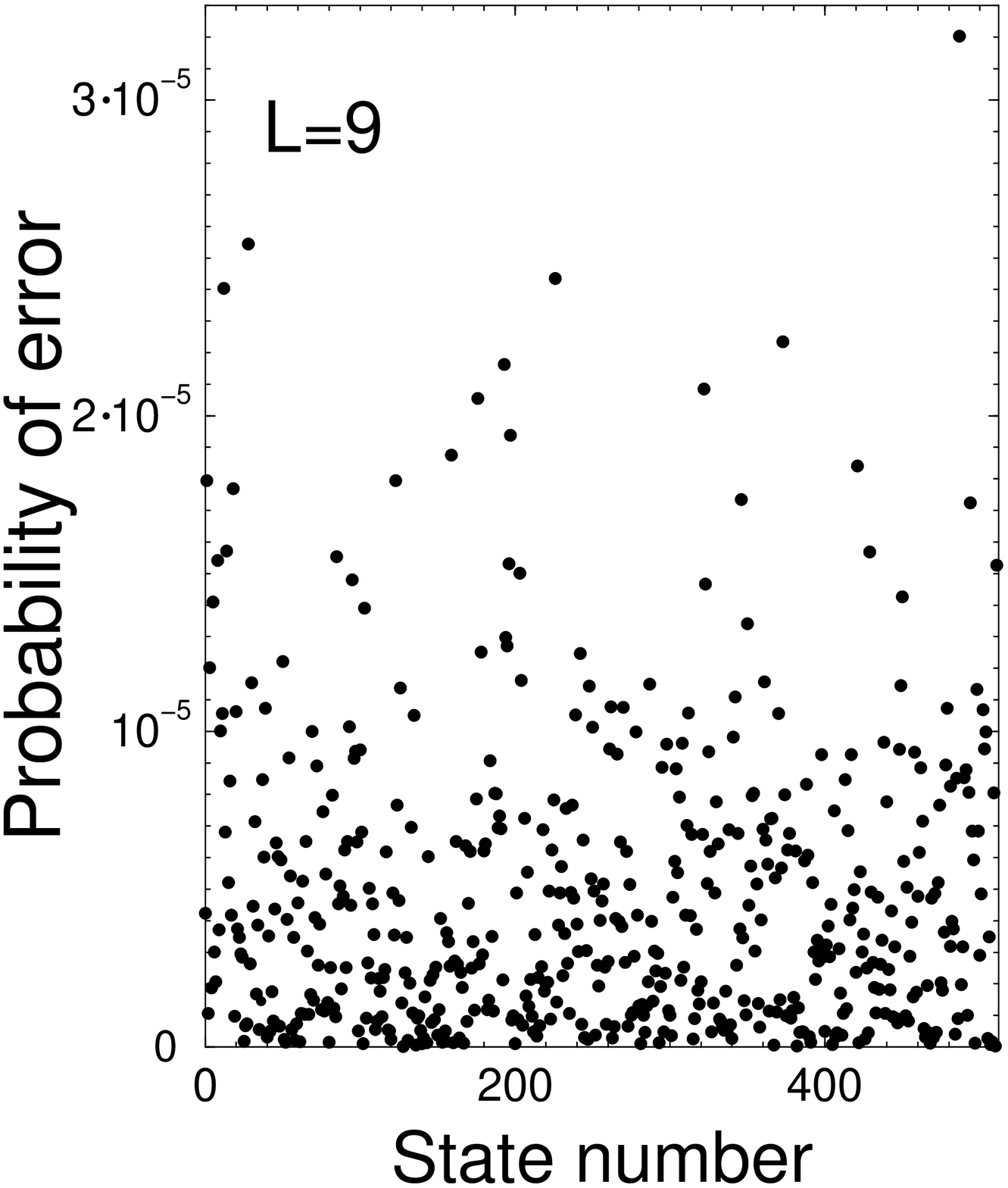}
\includegraphics[width=8cm,height=6cm]{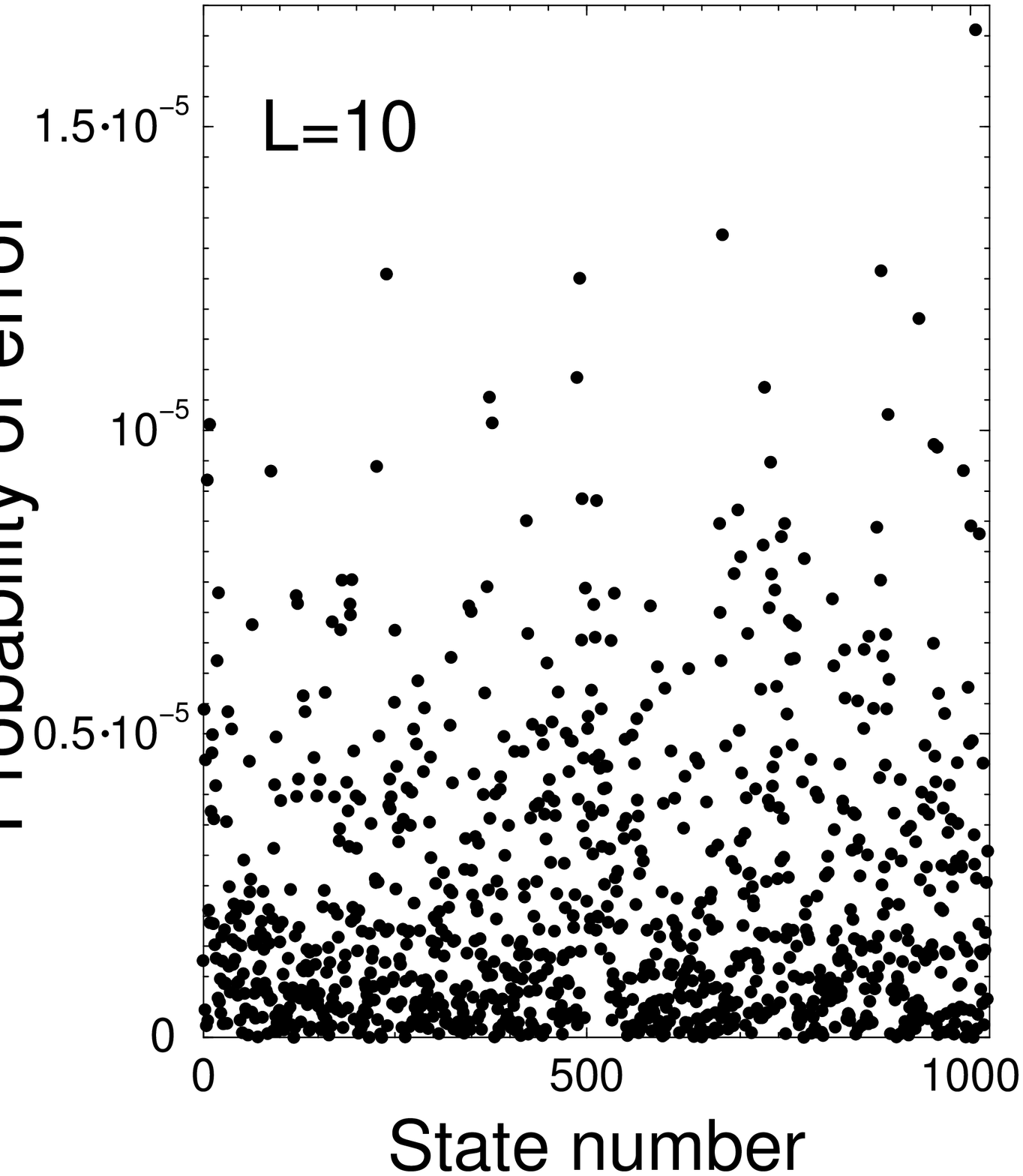}}
\vspace{-7mm}
\caption{The probability of errors $P_j(L)$ for different states of the
superposition and for different values of $L$.
$\delta\omega=10^4$, $\Omega=\Omega_k$ [see Eq. (\ref{Omegak})],
$k=2$, $J=1$.}
\label{fig:3}
\end{figure}

In Fig. 2 we plot the phase error as a function of time.
The phase error is defined as a the maximum
max$|\varphi_j-\Phi|$, $j=0,\dots,2^L-1$.
In Figs. 1(a,b) the phase error is equal to the distance between
the two dashed lines. During implementation of
each elementary CN gate between the neighboring qubits
(using 12 pulses for intermediate qubits) the
phases of different states of the superposition are different,
but after implementation of each gate CN$_{i,i+1}$ or CN$_{i+1,i}$
the phases are equalized, since our CN gates
between the neighboring qubits are probability and phase-corrected.
That is why in Fig. 2 the phase error is plotted with the
interval of 12 pulses
for intermediate qubits and 9 or 10 pulses for the edge qubits.
As follows from the results of the calculations,
the phase error is mostly generated by the pulses $Q_j^{00}$ and
$Q_j^{11}$.
From Fig. 2 one can see that the phase error grows linearly with
the number of these pulses. For larger number of qubits in the chain the
error is larger, because the implementation of the protocol requires
larger number of pulses.
[The total number of pulses $Q^{mn}_j$ is $2\times 36\times(L-2)-5$.]

The probability errors for the gate CN$_{0,L-1}$ are
defined as the moduli of the difference
$P_j(L)=\left||B_j|^2-|{B'}_{j}|^2\right|$,
where the coefficients $B_j$ and ${B'}_{j}$ in
Eq. (\ref{supperposition2}) are computed for the systems of
$L$ qubits. The values of $P_j(L)$
are shown in
in Fig. 3, and the relative errors
$P_j(L)/|B_j|^2$, are
plotted in Fig. 4 for different states of the
superposition.
From Fig. 3 one can see that the absolute
values of probability errors $P_j(L)$
decrease with $L$ increasing, since the initial
occupation probabilities are smaller for larger values of $L$
(because of the normalization condition $\sum_{j=0}^{2^L-1}|B_j|^2=1$).
On the other hand, as Fig. 4 demonstrates, the relative probabilities of error increase
with $L$ increasing because of increasing the number of pulses in the
protocol.

\begin{figure}
\mbox{\includegraphics[width=8cm,height=6cm]{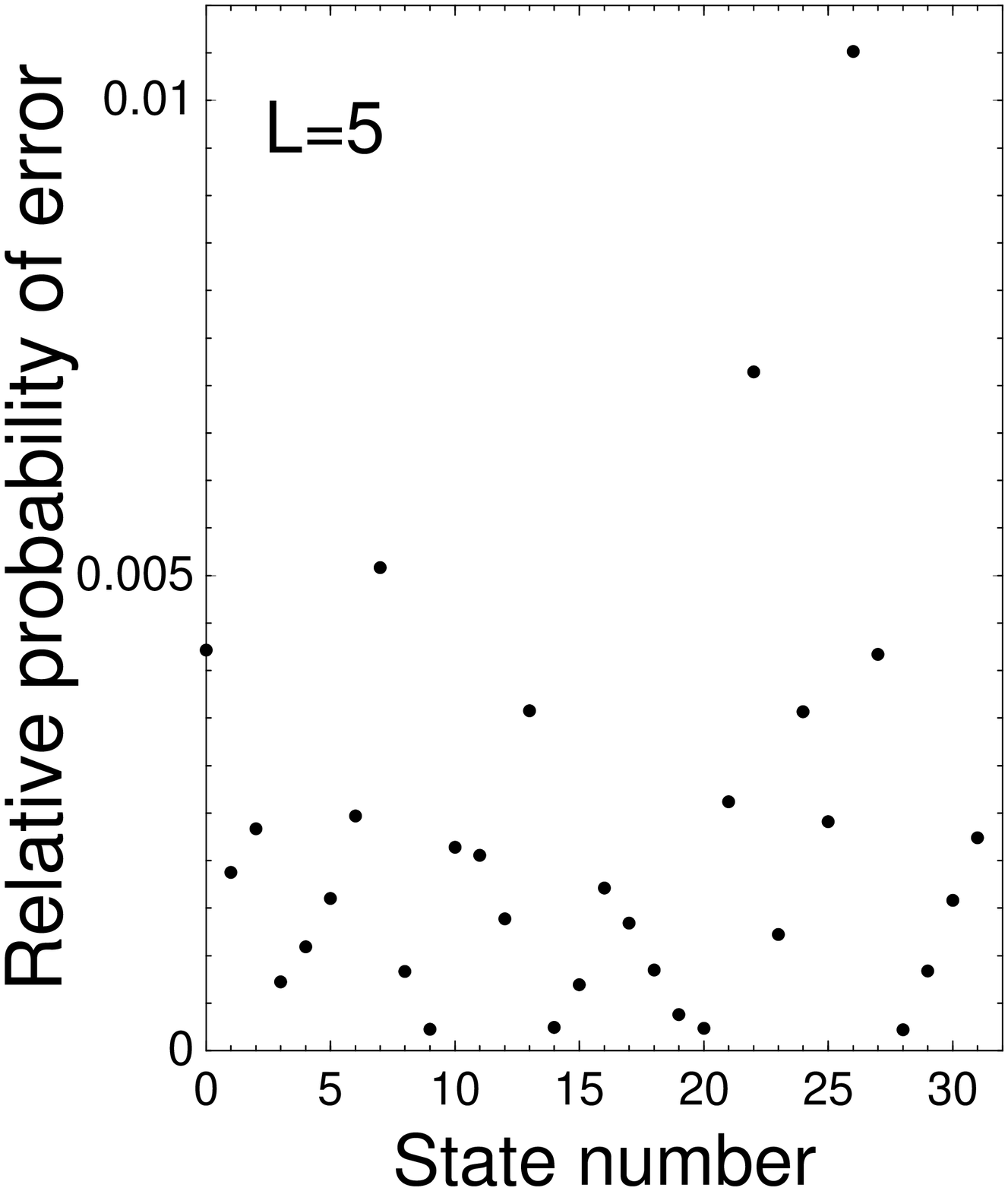}~
\includegraphics[width=8cm,height=6cm]{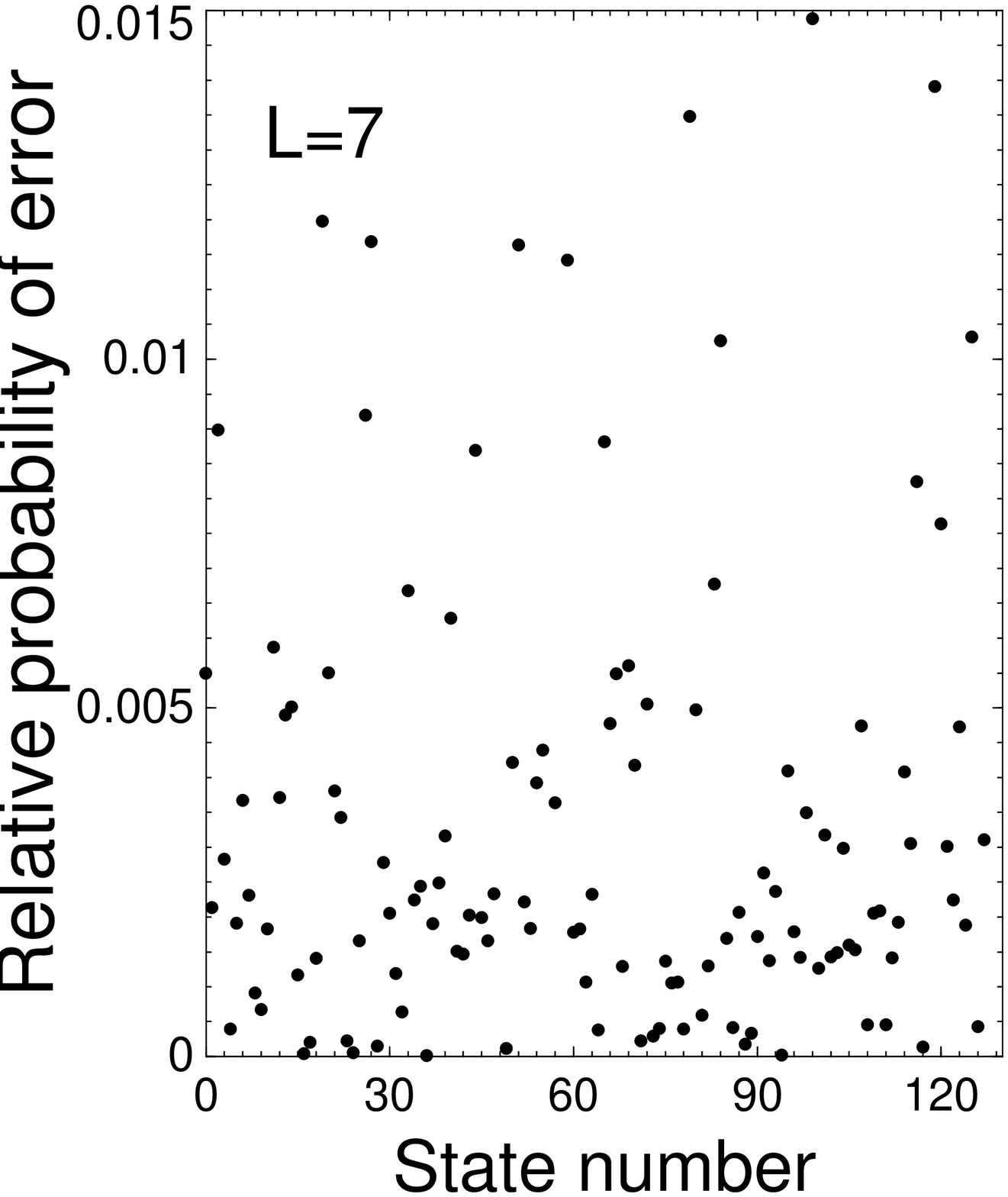}}
%\vspace{-10mm}
\mbox{\includegraphics[width=8cm,height=6cm]{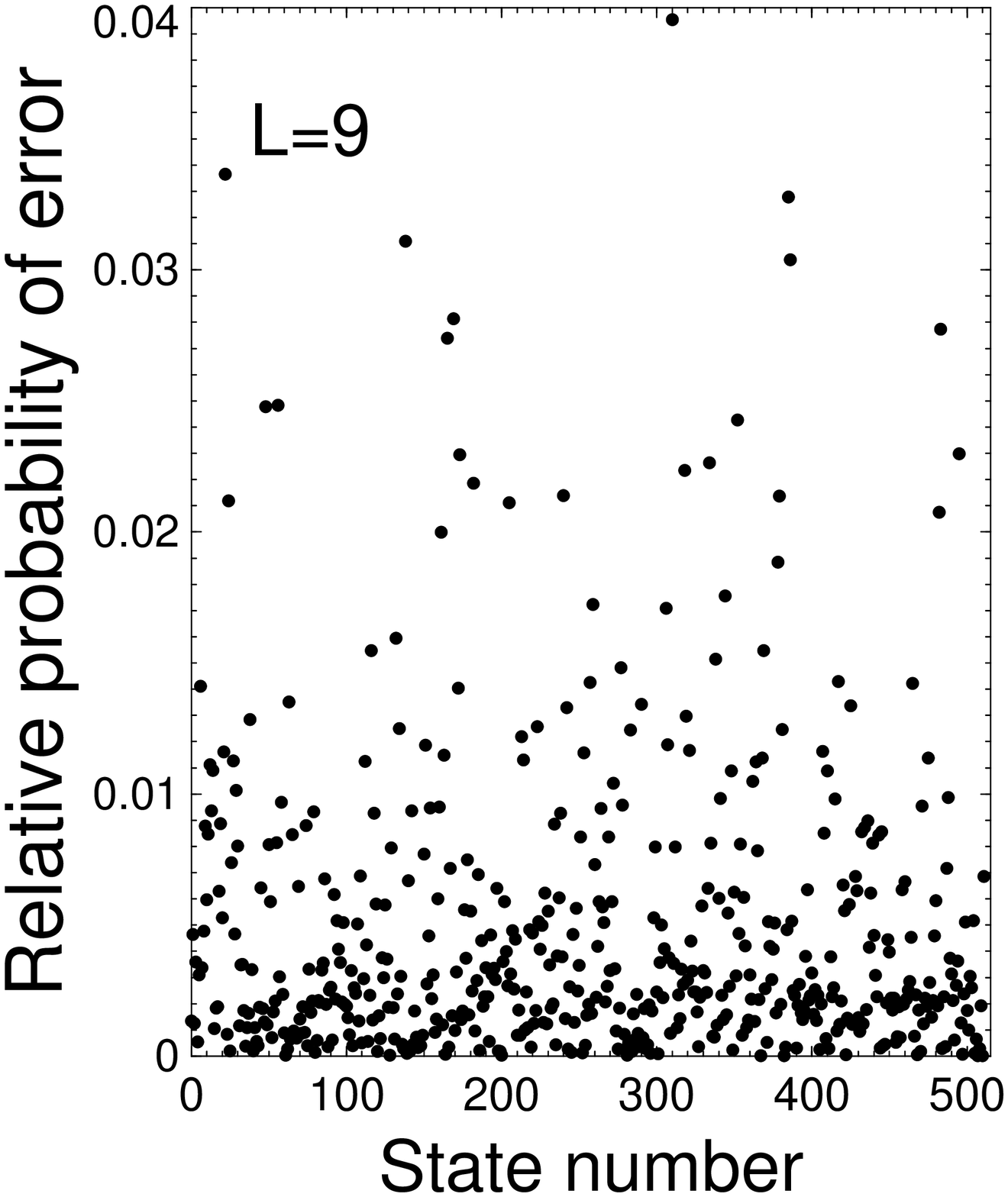}
\includegraphics[width=8cm,height=6cm]{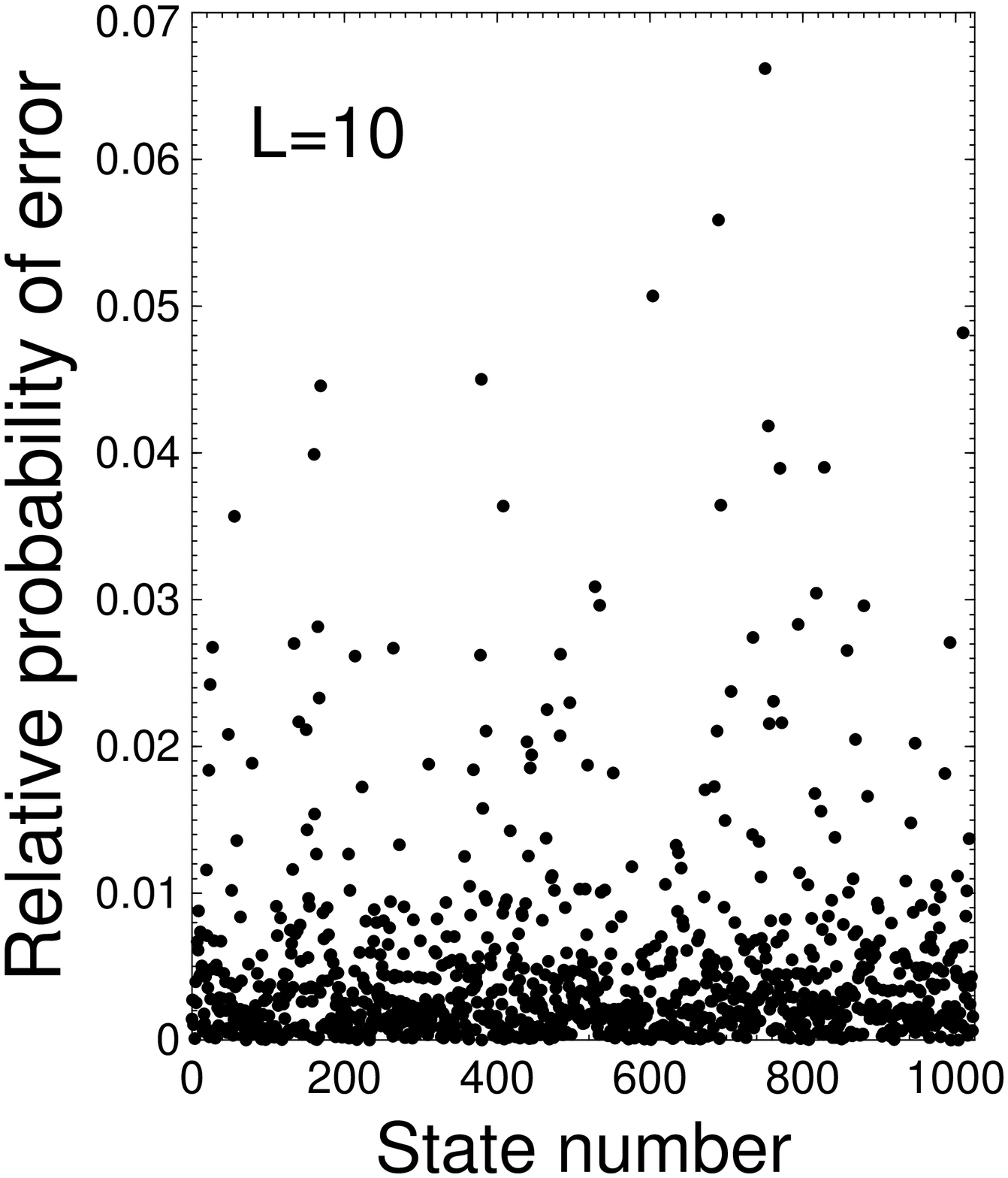}}
\vspace{-7mm}
\caption{The relative probability of errors $P_j(L)/|B_j|^2$
for different states of the
superposition and for different values of $L$. Other parameters
are the same as in Fig. 3.}
\label{fig:4}
\end{figure}

\section{Conclusion}
In this paper the first attempt is made to develop the
protocols for implementation of the universal quantum gates
on an arbitrary superposition of quantum states
in a multi-spin quantum computer.
The quantum gates
are constructed for a homogeneous spin chain
(the spins and interactions between them are identical).
In spite of the fact that such a computer is not realized experimentally
(because of severe technological challenges),
it provides a relatively simple
model with basic features common for many solid-state
scalable quantum computer proposals.

We solved the general problems which one should face
working with superpositional states in
different types of quantum computers characterized by the
constant (not switchable) interaction between qubits
(for example in Kane's quantum computer ~\cite{Kane}).
Since the transition frequencies required to flip the $j$th qubit in the
states $|\dots0_{j+1}q_j0_{j-1}\dots\rangle$,
$|\dots 1_{j+1}q_j0_{j-1}\dots\rangle$
(or $|\dots0_{j+1}q_j1_{j-1}\dots\rangle$), and
$|\dots1_{j+1}q_j1_{j-1}\dots\rangle$ ($q_j=0,1$) are different, one
inevitably will create an error by flipping the $j$th qubit.
For an Ising spin quantum computer we corrected this error using
additional correcting pulse which removes unwanted states from the
register of quantum computer,
created in the result of the unwanted transitions.

Another important problem solved in this paper is a minimization of the
phase errors. Even in the situation, when the probability errors are
minimized, each pulse of the protocol produces different
phases for different quantum states of the superposition.
In the Ising spin quantum computer the phases can be equalized
by proper choice of phases of the {\it rf} pulses.
Each protocol (for CN gate, one-qubit rotation, and others)
has its own set of phases.

The phase and probability errors studied in this paper, decrease
when the Rabi frequency, $\Omega$, decreases or the
frequency difference between the neighboring qubits, $\delta\omega$,
increases. In particular, all errors tend to zero when
$\Omega\rightarrow 0$ or $\delta\omega\rightarrow\infty$.

Our simulations show that the phase error less than 0.1 radians
for the studied protocols can be
achieved for the ratio $\delta\omega/J>10^4$. In our example with
phosphorus impurity donors in silicon (see Introduction) the ratio
$\delta\omega/J$ is about 170. We will mention three ways to increase
this ratio. The first way is to increase $\delta\omega$ using even
greater gradients of the magnetic field suggested in~\cite{6}.
The second
way is to decrease the interaction constant $J$.
This can be easily achieved by
either increase of the distance between the impurity atoms or using the
atoms with smaller effective radius. However decrease of $J$ will reduce
the clock speed of a quantum computer below MHz region. The third way
is to develop more sophisticated sequences of electromagnetic pulses,
which further suppress the phase error caused by non-resonant
transitions.

\begin{acknowledgments}
We are thankful to T. Seligman for useful discussions.
This work was supported by the Department of Energy (DOE) under
Contract No. W-7405-ENG-36, by the National Security Agency (NSA),
and by the
Advanced Research and Development Activity (ARDA).
The work of C.P. was supported by
Direcci\'on General de Estudios de Posgrado (DGEP).
C.P. is thankful to the Center for Nonlinear Studies at the Los
Alamos National Laboratory for hospitality.
\end{acknowledgments}

\appendix
\section{Correcting pulse}
\label{sec:appendixA}
Suppose that only the state $|m\rangle=|\dots 1_{i+1}0_i0_{i-1}\rangle$
is initially populated, $C_m(t_0)=1$. We apply the pulse
$P_{i}^{11}$ with
the frequency $\nu_{i}^{11}=\omega_i-2J$, where $\omega_i$ is the
Larmor frequency of the $i$th qubit. The detuning for the
transition
$|m\rangle\rightarrow|p\rangle=|\dots 1_{i+1}1_i0_{i-1}\rangle$ is
$\Delta_{pm}=E_p-E_m-\nu_{i}^{11}=\Delta$, where $\Delta=2J$. From
Eq. (\ref{2x2}) after the pulse with the frequency $\nu_i^{11}$
and phase $\Phi$ we have
\begin{equation}
\label{aIn}
C_m=[\cos\alpha(k)+if(k)\sin\alpha(k)]e^{-i\theta(k)/2}
\end{equation}
$$
C_p=ig(k)\sin\alpha(k) e^{i[\theta(k)/2-\Phi+\Delta t_0]}.
$$
The Rabi frequency $\Omega_2$ of this pulse
is chosen to suppress the transitions $T_i^{00}$. In this case
the quantities $\alpha$, $\theta$, $f$, and $g$, defined below,
are the functions of only one parameter $k$. From the
$2\pi k$-condition one has
\begin{equation}
\label{a2pik2}
{\Delta_2\over\Omega_2}=\sqrt{4k_2^2-1},\qquad \Delta_2=2\Delta.
\end{equation}
We choose
\begin{equation}
\label{a2pik2a}
\Omega_2=2\Omega,\qquad k_2=k.
\end{equation}

For the transition $T_i^{10}$ one has $\Delta_{pm}=\Delta$
in Eq. (\ref{aIn}), so that
\begin{equation}
\label{alpha}
\alpha(k)=\frac{\tau_2} 2\sqrt{\Delta^2+\Omega_2^2}=
\frac\pi 2\sqrt{{\left(\Delta\over\Omega_2\right)^2+1}}=
\frac\pi 2\sqrt{k^2+3/4},
\end{equation}
where $\tau_2=\pi/\Omega_2$ is the duration of the first pulse.
The other quantities are
\begin{equation}
\label{psi}
\theta(k)=\Delta\tau_2={\Delta\pi\over \Omega_2}=
\frac\pi 2\sqrt{4k^2-1},
\end{equation}
\begin{equation}
\label{f}
f(k)={\Delta\over \sqrt{\Delta^2+\Omega_2^2}}=
\sqrt{{k^2-1/4\over k^2+3/4}},
\end{equation}
\begin{equation}
\label{g}
g(k)={\Omega_2\over \sqrt{\Delta^2+\Omega_2^2}}=
{1\over\sqrt{k^2+3/4}}
\end{equation}

After the correcting pulse with the frequency $\nu_i^{10}$,
phase $\varphi_c^{11}$ (here the superscript indicates that the
first pulse has the form $P_i^{11}$),
and the duration $\tau_c$, one has
\begin{equation}
\label{correctIn}
C_m(t)=C_m\cos\beta+iC_p\sin\beta e^{i\varphi_c^{11}},
\end{equation}
$$
C_p(t)=C_p\cos\beta+iC_m\sin\beta e^{-i\varphi_c^{11}},
$$
where
\begin{equation}
\label{beta_tau}
t=t_0+\tau_c,\qquad \beta={\Omega_c\tau_c\over 2}.
\end{equation}
From the conditions ${\rm Re}\{C_p(t)\}=0$ and ${\rm Im}\{C_p(t)\}=0$,
where Re and Im stand for, respectively, the real and imaginary parts of
$C_p(t)$, one obtains the following equations:
\begin{equation}
\label{reIm}
\tan\Theta=-f\tan\alpha,\qquad
\varphi_c^{11}=-\theta+\Phi-\Delta t_0-\Theta,\qquad
\tan\beta=-g\tan\alpha\cos\Theta,
\end{equation}
where we do not indicate dependence of parameters on $k$.
The second and third equations in (\ref{reIm}) define, respectively,
the phase $\varphi_c^{11}$ and the duration
of the correcting pulse required to correct the error.

If one considers the transitions generated by the pulse $P_i^{11}$
in the opposite direction
$|\dots 1_{i+1}1_i0_{i-1}\dots\rangle\rightarrow
|\dots 1_{i+1}0_i0_{i-1}\dots\rangle$, one should just change
the signs as [see Eqs. (\ref{2x2}) and (\ref{2x2a})]
\begin{equation}
\label{signs}
f\rightarrow -f~~\theta\rightarrow -\theta,~~
\Phi\rightarrow-\Phi,~~\Delta t_0\rightarrow -\Delta t_0,~~
\varphi_c\rightarrow-\varphi_c
\end{equation}
in Eq. (\ref{reIm}). After this transformation both sides of the second
Eq. (\ref{reIm}) change their sign, so that the expressions for the
parameters of the correcting pulse, $\varphi_c^{11}$ and
$\beta$, do not change. Hence, one can
suppress both unwanted transitions,
$|\dots 1_{i+1}0_i0_{i-1}\dots\rangle\rightarrow
|\dots 1_{i+1}1_i0_{i-1}\dots\rangle$ and
$|\dots 1_{i+1}1_i0_{i-1}\dots\rangle\rightarrow
|\dots 1_{i+1}0_i0_{i-1}\dots\rangle$, simultaneously.

For the transitions generated by the pulse $P_i^{00}$ one should
change the sign of $\Delta$, so that the expression for the
phase $\varphi_c^{00}$ of the correcting pulse takes the form
\begin{equation}
\label{signs1}
\varphi_c^{00}=\theta+\Phi+\Delta t_0+\Theta,
\end{equation}
where $\Theta$ is defined in the first equation (\ref{reIm}).

For $k>1$ the value of $\beta$ is small and negative.
We take the duration of the pulse $\beta^*=\beta+\pi$, so that
$C_m(t)$ in Eq. (\ref{correctIn})
changes its sign, and $C_p(t)$ is again equal to zero.
After the correcting pulse
one has
\begin{equation}
\label{corrected}
C_m(t)=-\exp(-i\theta/2-i\Theta),\qquad C_p(t)=0.
\end{equation}

One can suppress the non-resonant transitions, $T^{00}_i$ and $T^{11}_i$
generated by the correcting pulse with moduli of detunings $\Delta$,
choosing the value of $\Omega_c$ satisfying
$2\pi k$-condition,
\begin{equation}
\label{2pik_c}
{\lambda_c\tau_c\over 2}=\pi k_c, \qquad
\lambda_c=\sqrt{\Delta^2+\Omega_c^2}, \qquad
\tau_c={2\beta^*\over\Omega_c},
\end{equation}
where $k_c$ is the integer number.
From Eq. (\ref{2pik_c}) one has
\begin{equation}
\label{deltaOmega_c}
{\Delta\over\Omega_c}=\sqrt{\left({\pi k_c\over\beta^*}\right)^2-1}.
\end{equation}
Below we take $k_c=k$. Since $\beta^*\approx\pi$,
one has $\Omega_c\approx2\Omega_1\approx\Omega_2$ where
$\Omega_1$ and $\Omega_2$ are defined in Eqs.
(\ref{deltaOmega}) and (\ref{deltaOmega2}).

\section{Unwanted phases}
\label{sec:appendixB}
We calculate here unwanted phases generated by the probability-corrected
pulses $Q_i^{01}(\varphi)$, $Q_i^{00}(\varphi)$, and $Q_i^{11}(\varphi)$.
From the first equation in (\ref{corrected}) the pulse $Q_i^{11}$ generates
the phase $\pi-\theta/2-\Theta$ for the states
$|\dots 1_{i+1}0_i0_{i-1}\dots\rangle$ and
$|\dots 0_{i+1}0_i1_{i-1}\dots\rangle$.
From Eq. (\ref{signs})
the same pulse generates the phases $\pi+\theta/2+\Theta$ for the states
$|\dots 1_{i+1}1_i0_{i-1}\dots\rangle$ and
$|\dots 0_{i+1}1_i1_{i-1}\dots\rangle$.
In a similar way one can deduce the phases generated by the pulse
$Q_i^{00}(\varphi)$ for the states $|\dots 1_{i+1}q_i0_{i-1}\dots\rangle$
and $|\dots 0_{i+1}q_i1_{i-1}\dots\rangle$, where $q_i=0,1$.

Now, based on the results of Appendix \ref{sec:appendixA},
we calculate the phase generated by the pulse $Q_i^{11}(\varphi)$
for the state
$|\dots 0_{i+1}0_i0_{i-1}\rangle\dots$. Since detuning for the transition
$|\dots 0_{i+1}0_i0_{i-1}\dots\rangle\rightarrow
|\dots 0_{i+1}1_i0_{i-1}\dots\rangle$ is positive and equal to
$\Delta_2=2\Delta$, from Eq. (\ref{2x2})
the first $\pi$ pulse [of the combined pulse $Q_i^{11}(\varphi)$]
generates the phase $-\Delta_2/\Omega_2=-\Delta/\Omega=-\theta$.
The duration of the correcting pulse with the frequency
$\nu_i^{10}$ is defined by
the third equation in Eqs. (\ref{2pik_c}), and the detuning is $\Delta$.
From Eq. (\ref{2x2}) the phase acquired by the state
$|\dots 0_{i+1}0_i0_{i-1}\rangle\dots$ in the result of action
of the correcting pulse is
\begin{equation}
\label{phase_c}
-{\Delta\tau_c\over 2}=-\beta^*{\Delta\over\Omega_c}=
-\sqrt{(\pi k_c)^2-\beta^{*2}}\equiv-\gamma,
\end{equation}
where we used Eq. (\ref{deltaOmega_c}).
After the action of two pulses, which constitute the probability corrected
pulse $Q_i^{11}(\varphi)$, the phase acquired by the state
$|\dots 0_{i+1}0_i0_{i-1}\dots\rangle$ is $-\theta-\gamma$.
We should note that for both pulses (main and correcting) the transition
$|\dots 0_{i+1}0_i0_{i-1}\dots\rangle\rightarrow
|\dots 0_{i+1}1_i0_{i-1}\dots\rangle$ is suppressed by the $2\pi k$
condition. In a similar way one can deduce other phases
for other pulses in the Table I.

\section{Protocol for the CN gate}
\label{sec:AppendixC}
Our aim is to implement the CN gate between
neighboring qubits, $CN_{a,b}$, using only operations which
affect the control and the target qubits, $q_a$ and $q_{b}$,
respectively.  Thus, for complete generality we need to consider
the states of the control and target qubits, as well as their
neighbors.
The CN gate in a homogeneous spin chain
without phase correction can be implemented by the following sequence
\begin{equation}\label{eq:CN1}
Q_a^{11}(\varphi_8)Q_a^{10}(\varphi_7)Q_a^{00}(\varphi_6)
Q_{b}^{00}(\varphi_5)Q_a^{11}(\varphi_4)Q_a^{10}(\varphi_3)
Q_a^{00}(\varphi_2)Q_{b}^{11}(\varphi_1).
\end{equation}
This sequence can be described as follows:
(in terms of the transformations for the amplitudes,
we note that the phases of the various states will not be equal)
(i) a controlled flip of
$q_{b}$ if both of its neighbors are in the ``1'' state, (ii) a
Not operation on $q_a$, (iii) a controlled flip of $q_{b}$ if
both of it's neighbors are in the ``0'' state, and finally (iv)
another Not operation on $q_a$ which returns it to its initial
state.

The transformation resulting from (\ref{eq:CN1}) is
(we indicate in this Appendix only four relevant qubits of the
spin chain),

{\setlength\arraycolsep{2pt}\begin{eqnarray}\label{eq:CNphase1}
  &\ &|0000\rangle\rightarrow e^{i[-\gamma-\theta/2+\Theta-\varphi_2+\varphi_6]}|0000\rangle,
  \nonumber \\
  &\ &|0001\rangle\rightarrow e^{i[+\gamma+\theta/2-\Theta-\varphi_2+\varphi_6]}|0001\rangle,
  \nonumber \\
  &\ &|0010\rangle\rightarrow e^{i[+\gamma+\theta/2-\Theta-\varphi_3+\varphi_7]}|0010\rangle,
  \nonumber \\
  &\ &|0011\rangle\rightarrow e^{i[-\gamma-\theta/2+\Theta-\varphi_3+\varphi_7]}|0011\rangle,
  \nonumber \\
  &\ &|0100\rangle\rightarrow e^{i[\pi/2+\gamma+3\theta/2-\Theta+\varphi_3+\varphi_5-\varphi_6]}
  |0110\rangle,
  \nonumber \\
  &\ &|0101\rangle\rightarrow e^{i[\pi/2+\gamma+\theta/2+\Theta+\varphi_1+\varphi_2-\varphi_6]}
  |0111\rangle,
  \nonumber \\
  &\ &|0110\rangle\rightarrow e^{i[\pi/2-\gamma-3\theta/2+\Theta+\varphi_2-\varphi_5-\varphi_7]}
  |0100\rangle,
  \nonumber \\
  &\ &|0111\rangle\rightarrow e^{i[\pi/2-\gamma-\theta/2-\Theta-\varphi_1+\varphi_3-\varphi_7]}
  |0101\rangle,
  \nonumber \\
  &\ &|1000\rangle\rightarrow e^{i[-\gamma-\theta/2+\Theta-\varphi_3+\varphi_7]}|1000\rangle,
  \nonumber \\
  &\ &|1001\rangle\rightarrow e^{i[+\gamma+\theta/2-\Theta-\varphi_3+\varphi_7]}|1001\rangle,
  \nonumber \\
  &\ &|1010\rangle\rightarrow e^{i[+\gamma+\theta/2-\Theta-\varphi_4+\varphi_8]}|1010\rangle,
  \nonumber \\
  &\ &|1011\rangle\rightarrow e^{i[-\gamma-\theta/2+\Theta-\varphi_4+\varphi_8]}|1011\rangle,
  \nonumber \\
  &\ &|1100\rangle\rightarrow e^{i[\pi/2-\gamma-\theta/2+3\Theta+\varphi_4+\varphi_5-\varphi_7]}
  |0110\rangle,
  \nonumber \\
  &\ &|1101\rangle\rightarrow e^{i[\pi/2+\gamma+\theta/2+\Theta+\varphi_1+\varphi_3-\varphi_7]}
  |1111\rangle,
  \nonumber \\
  &\ &|1110\rangle\rightarrow e^{i[\pi/2+\gamma+\theta/2-3\Theta+\varphi_3-\varphi_5-\varphi_8]}
  |1100\rangle,
  \nonumber \\
  &\ &|1111\rangle\rightarrow e^{i[\pi/2-\gamma-\theta/2-\Theta-\varphi_1+\varphi_4-\varphi_8]}
  |1101\rangle.
\end{eqnarray}}

However, when this sequence of gates is used, it is impossible to
make the phases on all of the states equal.  This can easily be
seen by examining, for example, the first two transformations in
Eq. (\ref{eq:CNphase1}).  Here, it's impossible to equalize the phases
on the two final states, for arbitrary values of $\theta$, $\Theta$,
and $\gamma$. Thus, in order
to produce a phase-corrected CN gate, it's necessary to add
extra pulses which can introduce more controllable phases
$\varphi_j$.

A sequence that we have found to work is,

{\setlength\arraycolsep{2pt}\begin{eqnarray}\label{eq:CN2a}
CN_{i,i+1}&=&Q_i^{11}(\varphi_{10})Q_i^{10}(\varphi_9)
Q_i^{00}(\varphi_8)\underbrace{Q_{i+1}^{01}(0)
Q_{i+1}^{01}(\varphi_7)}
Q_{i+1}^{00}(\varphi_6) \nonumber \\
& \ & Q_i^{11}(\varphi_5)Q_i^{10}(\varphi_4)
Q_i^{00}(\varphi_3)\underbrace{Q_{i+1}^{01}(0)Q_{i+1}^{01}(\varphi_2)}
Q_{i+1}^{11}(\varphi_1),
\end{eqnarray}}
where the braces indicate additional operations not found
in Eq.~(\ref{eq:CN1}). The transformation which results from this
sequence is,
{\setlength\arraycolsep{2pt}\begin{eqnarray}\label{eq:CNphase2}
  &\ &|0000\rangle\rightarrow e^{i[\pi-\gamma-5\theta/2+\Theta-\varphi_3-\varphi_7+\varphi_8]}
  |0000\rangle,
  \nonumber \\
  &\ &|0001\rangle\rightarrow e^{i[\pi+\gamma+5\theta/2-\Theta-\varphi_2-\varphi_3+\varphi_8]}
  |0001\rangle,
  \nonumber \\
  &\ &|0010\rangle\rightarrow e^{i[\pi+\gamma+5\theta/2-\Theta-\varphi_4+\varphi_7+\varphi_9]}
  |0010\rangle,
  \nonumber \\
  &\ &|0011\rangle\rightarrow e^{i[\pi-\gamma-5\theta/2+\Theta+\varphi_2-\varphi_4+\varphi_9]}
  |0011\rangle,
  \nonumber \\
  &\ &|0100\rangle\rightarrow e^{i[3\pi/2+\gamma-\theta/2+\varphi_2+\varphi_4+\varphi_6
  -\varphi_8]}|0110\rangle,
  \nonumber \\
  &\ &|0101\rangle\rightarrow e^{i[3\pi/2+\gamma+5\theta/2+\Theta+\varphi_1+\varphi_3-\varphi_7
  -\varphi_8]}|0111\rangle,
  \nonumber \\
  &\ &|0110\rangle\rightarrow e^{i[3\pi/2-\gamma+\theta/2+\Theta-\varphi_2+\varphi_3-\varphi_6
  -\varphi_9]}|0100\rangle,
  \nonumber \\
  &\ &|0111\rangle\rightarrow e^{i[3\pi/2-\gamma-5\theta/2-\Theta-\varphi_1+\varphi_4+\varphi_7
  -\varphi_9]}|0101\rangle,
  \nonumber \\
  &\ &|1000\rangle\rightarrow e^{i[\pi-\gamma-5\theta/2+\Theta-\varphi_4-\varphi_7+\varphi_9]}
  |1000\rangle,
  \nonumber \\
  &\ &|1001\rangle\rightarrow e^{i[\pi+\gamma+5\theta/2-\Theta-\varphi_2-\varphi_4+\varphi_9]}
  |1001\rangle,
  \nonumber \\
  &\ &|1010\rangle\rightarrow e^{i[\pi-\gamma-5\theta/2+\Theta-\varphi_5+\varphi_7+\varphi_{10}]}
  |1010\rangle,
  \nonumber \\
  &\ &|1011\rangle\rightarrow e^{i[\pi+\gamma+5\theta/2-\Theta+\varphi_2-\varphi_5+\varphi_{10}]}
  |1011\rangle,
  \nonumber \\
  &\ &|1100\rangle\rightarrow e^{i[3\pi/2-\gamma-5\theta/2+3\Theta+\varphi_2+\varphi_5+\varphi_6
  -\varphi_9]}|0110\rangle,
  \nonumber \\
  &\ &|1101\rangle\rightarrow e^{i[3\pi/2+\gamma+5\theta/2+\Theta+\varphi_1+\varphi_4-\varphi_7
  -\varphi_9]}|1111\rangle,
  \nonumber \\
  &\ &|1110\rangle\rightarrow e^{i[3\pi/2+\gamma+5\theta/2-3\Theta-\varphi_2+\varphi_4-\varphi_6
  -\varphi_{10}]}|1100\rangle,
  \nonumber \\
  &\ &|1111\rangle\rightarrow e^{i[3\pi/2-\gamma-5\theta/2-\Theta-\varphi_1+\varphi_5+\varphi_7
  -\varphi_{10}]}|1101\rangle.
\end{eqnarray}}

This transformation introduces more than enough independent
variables, {$\varphi_j$}, to equalize all phases.  To find the
correct values for the $\varphi_j$'s we set all the phases
above equal to the same number, and solve the resulting
system of equations.  The solution in the form

{\setlength\arraycolsep{2pt}\begin{eqnarray}
\label{c5}
 &
 \varphi_1=-2\gamma-5\theta,\quad
 \varphi_2=\gamma+\frac{5\theta}{2}-\Theta,\quad
 \varphi_3=\frac{3\pi}{4}+2\gamma+2\theta-4\Theta+2\varphi_9-\varphi_{10},\quad
 & \nonumber \\ &
 \varphi_4=\frac{3\pi}{4}+\varphi_9,\quad
 \varphi_5=\frac{3\pi}{4}+\varphi_{10},\quad
 \varphi_6=-2\Theta+\varphi_9-\varphi_{10},\quad
 \varphi_7=-\gamma-\frac{5\theta}{2}+\Theta,\quad
 & \nonumber \\ &
 \varphi_8=2\gamma+2\theta-4\Theta+2\varphi_9-\varphi_{10}, &
\end{eqnarray}}
equalizes the phases of all states to $\pi/4$.
There are only eight independent $\varphi_j$'s in these equations,
so that $\varphi_9$ and $\varphi_{10}$ can be set to zero. Thus,
the pulse sequence (\ref{eq:CN2a}) with the phases (\ref{c5})
implements a phase corrected CN gate.

{}

\end{document}